\documentclass{SciPost}

\usepackage[english]{babel}

\hypersetup{
    colorlinks,
    linkcolor={red!50!black},
    citecolor={blue!50!black},
    urlcolor={blue!80!black},
    pdftitle={MPS overlaps and Sklyanin determinants},
    pdfauthor={T. Gombor}
}

\usepackage[bitstream-charter]{mathdesign}
\DeclareSymbolFont{usualmathcal}{OMS}{cmsy}{m}{n}
\DeclareSymbolFontAlphabet{\mathcal}{usualmathcal}

\fancypagestyle{SPstyle}{
\fancyhf{}
\lhead{\colorbox{scipostblue}{\bf \color{white} ~SciPost Physics }}
\rhead{{\bf \color{scipostdeepblue} ~Submission }}

\fancyfoot[C]{\textbf{\thepage}}
}

\begin{document}

\pagestyle{SPstyle}

\begin{center}{\Large \textbf{\color{scipostdeepblue}{
MPS overlaps and Sklyanin determinants\\
}}}\end{center}

\begin{center}\textbf{
Tamas  Gombor\textsuperscript{1$\star$},
}\end{center}

\begin{center}
{\bf 1} HUN-REN Wigner Research Centre for Physics, Budapest, Hungary
\\[\baselineskip]
$\star$ \href{mailto:gombor.tamas@wigner.hun-ren.hu}{\small gombor.tamas@wigner.hun-ren.hu}
\end{center}

\section*{\color{scipostdeepblue}{Abstract}}
\textbf{\boldmath{%
We derive explicit overlap functions for integrable matrix product states associated with arbitrary finite-dimensional irreducible evaluation representations of the orthogonal and symplectic twisted Yangians. The central step is to relate the commuting quasi-determinants entering the universal overlap formula to Sklyanin determinants of nested subalgebras. For orthogonal evaluation representations, this relation determines their spectra directly in the Gelfand--Tsetlin basis. In the symplectic case, we combine Sklyanin determinants and minors with a factorization of the nested rank-two boundary matrices to obtain the corresponding spectra. The resulting overlap functions are explicit products parametrized by Gelfand--Tsetlin patterns and require only the highest-weight data of the underlying orthogonal or symplectic representation. Together with the universal ratio of Gaudin determinants and the model-dependent pseudovacuum overlap coefficients, they determine the normalized MPS overlaps. We also express the overlap functions in the orthogonal case beyond evaluation representations in terms of the highest weights arising along the nested twisted-Yangian subalgebras.
}}

\vspace{\baselineskip}



\vspace{10pt}
\noindent\rule{\textwidth}{1pt}
\tableofcontents
\noindent\rule{\textwidth}{1pt}
\vspace{10pt}


\section{Introduction}\label{sec:introduction}

Integrable quantum systems provide a setting in which interactions, correlations, and non-equilibrium dynamics can be studied by exact methods. Their extensive families of conserved quantities and Bethe-ansatz description make them valuable both as models of many-body physics and as benchmarks for more general approaches. Integrable matrix product states (MPSs) extend this framework from the spectral problem to the choice of boundary or initial state: they describe correlated states with a finite-dimensional internal space while retaining the algebraic constraints needed for exact overlap calculations \cite{Piroli:2017sei,Pozsgay:2018dzs}. In a quantum quench, these overlaps determine the expansion of the initial state in eigenstates of the post-quench Hamiltonian and provide essential input for the Quench Action description of stationary states \cite{Caux:2013ra,Essler:2016ufo,DeNardis:2014BEC,Brockmann:2014QuenchAction,Mestyan:2017NestedQuench,Bertini:2017HubbardQuench,Rylands:2022naf,Rylands:2022gev} and relaxation \cite{DeNardis:2015Relaxation}. They also enter calculations of Loschmidt echoes \cite{Piroli:2017Loschmidt} and diagonal R\'enyi entropies \cite{Alba:2017Renyi}. In defect versions of the AdS/CFT correspondence, tree-level one-point functions can likewise be expressed as overlaps of Bethe eigenstates with MPSs encoding the defect data \cite{deLeeuw:2015hxa,Buhl-Mortensen:2015gfd,deLeeuw:2016umh,DeLeeuw:2018cal,deLeeuw:2019ebw,Gombor:2022aqj,Kristjansen:2023ysz,Gombor:2024zru,Holguin:2025bfe,Chalabi:2025nbg,Coronado:2025xwk}. Exact MPS overlaps therefore connect non-equilibrium many-body physics with correlation functions in gauge theories with defects.

Following the N\'eel--XXZ determinant formula \cite{Brockmann_2014}, analogous MPS and nested-chain formulas were proposed and tested \cite{deLeeuw:2015hxa,Buhl-Mortensen:2015gfd,deLeeuw:2016umh,Piroli:2018ksf}. Systematic proofs followed from coordinate Bethe ansatz \cite{Jiang:2020sdw} and off-shell recursions in algebraic Bethe ansatz \cite{Gombor:2021uxz,Gombor:2021hmj,Gombor:2023bez}. A key advantage of the algebraic approach is its independence from the representations in the quantum space and its applicability to broad classes of boundary states, yielding general results beyond individual examples. The universal MPS formula was subsequently announced in \cite{Gombor:2024iix} and proved for a broad class of boundary representations using the KT relation in \cite{Gombor:2025wvu}.

In the universal formula, the normalized overlap is expressed as the product of a Gaudin-determinant ratio and an MPS-dependent sum. The determinant ratio depends only on the Bethe state, while the sum runs over the internal MPS space. Each internal degree of freedom contributes a scalar $\beta_{j}$ and $N-1$ functions $\tilde{\mathcal{F}}_{j}^{(k)}(u)$, with $k=1,\dots,N-1$. These coefficients and functions encode the boundary-state data independently of the chosen Bethe state. Their computation is therefore the remaining task in applying the universal formula.

Each integrable MPS is associated with a $K$-matrix satisfying the $\mathfrak{gl}_{N}$ boundary Yang--Baxter equation. From this matrix one constructs $N$ commuting operators $\mathbf{G}^{(k)}(u)$, with $k=1,\dots,N$, acting on the internal MPS space. Their construction involves quasi-determinants, with a modification in the symplectic case described below. They define a further family of commuting operators, $\mathbf{F}^{(k)}(u)=\left(\mathbf{G}^{(k)}(u)\right)^{-1}\mathbf{G}^{(k+1)}(u)$. After a spectral reparametrization and a scalar normalization, their eigenvalues give the functions $\tilde{\mathcal{F}}_{j}^{(k)}(u)$. A further operator $\mathbf{B}$, whose entries are the pseudovacuum overlaps of the untraced MPS, commutes with the F-operators. Its eigenvalues are the coefficients $\beta_{j}$
\footnote{This simple argument applies directly only for the $\mathfrak{so}_N$ symmetric MPSs. In the $\mathfrak{sp}_N$ case further deformations are necessary to obtain the coefficients $\beta_{j}$, see the details in \cite{Gombor:2025wvu}}.

The overlap problem thus reduces to constructing the quasi-determinants and determining their spectra. This can be done directly when explicit MPS matrices are available. For a family of states specified through representation theory, however, it is more useful to express the answer in terms of representation labels, as in the familiar example of a spin-$s$ representation of $\mathfrak{su}_{2}$. Here we pursue this approach for states associated with evaluation representations. Schematically, their boundary Yang--Baxter solutions take the form
\begin{equation}
\mathbf{K}_{i,j}(u)=\delta_{i,j}+\frac{1}{u\pm1/2}F_{i,j}.
\end{equation}
Here $F_{i,j}$ are generators in a representation of $\mathfrak{so}_{N}$ or $\mathfrak{sp}_{2n}$; the precise conventions are given below. To determine the overlaps, one needs both $\beta_{j}$ and $\tilde{\mathcal{F}}_{j}^{(k)}(u)$. The scalar coefficients depend not only on the $K$-matrix, but also on the $\mathfrak{gl}_{N}$ representations defining the quantum space, the chain length, and the normalization of the MPS. Since the overlap depends linearly on these coefficients, they can be determined separately, for example numerically. \emph{We focus on the functions $\tilde{\mathcal{F}}_{j}^{(k)}(u)$, which depend only on the $K$-matrix.}

To determine these functions, we use the Gelfand--Tsetlin (GT) basis to describe the spectra of the F-operators. Its vectors are labelled by patterns $\Lambda$ of weights $\lambda_{i,j}$, with admissibility conditions fixed by the highest weight of the underlying representation. As we show below, the overlap function associated with a pattern $\Lambda$ has the unified form
\begin{equation}
\tilde{\mathcal{F}}_{\Lambda}^{(k)}(u)=\Upsilon_{k}(u)\frac{\prod_{j=1}^{\left|\frac{k-1}{2}\right|_{\pm}}(u^{2}+(\gamma_{k-1,j}+1/2)^{2})}{\prod_{j=1}^{\left|k/2\right|_{\pm}}(u^{2}+\gamma_{k,j}^{2})}\frac{\prod_{j=1}^{\left|\frac{k+1}{2}\right|_{\pm}}(u^{2}+(\gamma_{k+1,j}+1/2)^{2})}{\prod_{j=1}^{\left|k/2\right|_{\pm}}(u^{2}+(\gamma_{k,j}+1)^{2})},
\end{equation}
where
\begin{equation}
\gamma_{k,j}=\lambda_{k,j}+\frac{k}{2}-j\mp\frac{1}{2}
\end{equation}
and
\begin{equation}
\Upsilon_{k}(u)=\begin{cases}
\frac{1}{\sqrt{u^{2}(u^{2}+1/4)}}, & \text{for }\mathfrak{so}_{N},\:k\text{ is odd},\\
\sqrt{u^{2}(u^{2}+1/4)}, & \text{for }\mathfrak{so}_{N},\:k\text{ is even},\\
-\sqrt{\frac{u^{2}}{u^{2}+1/4}}, & \text{for }\mathfrak{sp}_{2n}.
\end{cases}
\end{equation}
The upper and lower choices refer to the orthogonal and symplectic cases, respectively; the corresponding integer-part prescriptions are the floor and ceiling. The overlap is obtained by summing over all admissible GT patterns
\footnote{The definition of the Gaudin determinants $\det G^{\pm}$ can be found in several places e.g. in \cite{Gombor:2023bez}}:
\begin{equation}
\frac{\langle\mathrm{MPS}|\bar{u}\rangle}{\sqrt{\langle\bar{u}|\bar{u}\rangle}}=\sum_{\Lambda}\beta_{\Lambda}\prod_{k=1}^{N-1}\prod_{j=1}^{n_{k}/2}\tilde{\mathcal{F}}_{\Lambda}^{(k)}(u_{j}^{k})\times\sqrt{\frac{\det G^{+}}{\det G^{-}}}.\label{eq:overlap}
\end{equation}
The paper is organized as follows. Section~\ref{sec:preliminaries} fixes our Lie-algebra and highest-weight conventions, reviews GT patterns, and introduces the spin-chain, boundary-state, and twisted-Yangian constructions needed in the derivation. Section~\ref{sec:orthogonal} relates the orthogonal G-operators to nested Sklyanin determinants, derives the evaluation-representation overlap functions, and discusses the extension to general highest-weight data. Section~\ref{sec:symplectic} develops the modified symplectic construction using rank-two boundary matrices and their action on multiplicity spaces, leading to the corresponding GT-parametrized functions. Section~\ref{sec:conclusions} summarizes the results and outlines further directions. Appendix~\ref{app:conventions} relates the different $K$-matrix conventions, while Appendix~\ref{app:factorization} proves the determinant factorizations used in the main text.

\section{Preliminaries}\label{sec:preliminaries}

We begin by fixing the representation-theoretic conventions and the spin-chain notation used throughout the paper. After introducing the orthogonal and symplectic generators and their highest weights, we review the GT patterns that label the overlap functions. We then recall the Yangian description of Bethe states and the KT characterization of integrable MPSs. Finally, we introduce the Sklyanin minors and determinants that connect these boundary-state constructions to highest-weight data.

\subsection{Representations of the $\mathfrak{so}_{N}$ and $\mathfrak{sp}_{N}$ Lie algebras\label{sec:Representations-of-the}}

We use the following conventions for the generators of $\mathfrak{so}_{N}$:
\begin{equation}
\begin{split}F_{i,j} & =-F_{j,i},\\
\left[F_{i,j},F_{k,l}\right] & =\delta_{j,k}F_{i,l}-\delta_{i,l}F_{k,j}-\delta_{i,k}F_{j,l}+\delta_{j,l}F_{k,i}.
\end{split}
\end{equation}
For the symplectic Lie algebra, we use
\begin{equation}
\begin{split}F_{i,j} & =-(-1)^{i+j}F_{\bar{j},\bar{i}},\\
\left[F_{i,j},F_{k,l}\right] & =\delta_{j,k}F_{i,l}-\delta_{i,l}F_{k,j}-(-1)^{i+j}\delta_{\bar{i},k}F_{\bar{j},l}+(-1)^{i+j}\delta_{\bar{j},l}F_{k,\bar{i}}.
\end{split}
\end{equation}
In the $\mathfrak{sp}_{N}$ case, the indices satisfy $i,j,k,l=1,\dots,N$, with
\begin{equation}
\begin{split}\overline{2i} & =2i-1,\\
\overline{2i-1} & =2i.
\end{split}
\end{equation}
To define highest weights, we pass from these generators to a Cartan--Weyl basis.

The change of basis makes the raising operators and Cartan generators explicit and allows us to apply standard highest-weight formulas. Introduce the generator-valued matrix
\begin{equation}
F=\sum_{i,j=1}^{N}e_{i,j}\otimes F_{i,j},
\end{equation}
and the change-of-basis matrix
\begingroup\fontsize{8.2}{10.2}\selectfont
\begin{equation}
B=\begin{cases}
\frac{1}{\sqrt{2}}\sum_{j=1}^{n}(e_{2j-1,j}+e_{2j-1,N+1-j}+ie_{2j,j}-ie_{2j,N+1-j})+\delta_{N,2n+1}e_{2n+1,n+1}, & \mathfrak{so}_{N},\\
\sum_{j=1}^{m}(e_{2j-1,2j-1}+e_{2j,N-2j+2}+e_{N-2j+1,2j}+e_{N-2j+2,N-2j+1}), & \mathfrak{sp}_{4m},\\
\sum_{j=1}^{m}(e_{2j-1,2j-1}+e_{2j,N-2j+2}+e_{N-2j+1,2j}+e_{N-2j+2,N-2j+1})+e_{2m+1,2m+1}+e_{2m+2,2m+2}, & \mathfrak{sp}_{4m+2},
\end{cases}
\end{equation}
\endgroup
which satisfies
\begin{equation}
\begin{split}B^{T}B & =V,\quad\text{for }\mathfrak{so}_{N},\\
B^{T}UB & =V,\quad\text{for }\mathfrak{sp}_{N}.
\end{split}
\label{eq:BBV}
\end{equation}
Here $V$ is the $N\times N$ matrix with entries
\begin{equation}
V_{i,j}=\begin{cases}
\delta_{N+1-i,j}, & \text{for }\mathfrak{so}_{N},\\
(-1)^{j>n}\delta_{N+1-i,j}, & \text{for }\mathfrak{sp}_{N}.
\end{cases}\label{eq:Vdef}
\end{equation}
The transformed generators are defined by
\begin{equation}
F^{c}=B^{-1}FB=\sum_{i,j=1}^{N}e_{i,j}\otimes F_{i,j}^{c},
\end{equation}
and they satisfy
\begin{equation}
\begin{split}
F_{i,j}^{c} &= -\sigma_{i}\sigma_{j}F_{N+1-j,N+1-i}^{c},\\
\left[F_{i,j}^{c},F_{k,l}^{c}\right]  &= \delta_{j,k}F_{i,l}^{c}-\delta_{i,l}F_{k,j}^{c}-\sigma_{i}\sigma_{j}\delta_{N+1-i,k}F_{N+1-j,l}^{c}+\sigma_{i}\sigma_{j}\delta_{N+1-j,l}F_{k,N+1-i}^{c},
\end{split}
\end{equation}
where $\sigma_{i}=1$ for $\mathfrak{so}_N$ and 
\begin{equation}
\sigma_{i}=\begin{cases}
+1, & 1\leq i\leq n,\\
-1, & n<i\leq N,
\end{cases}
\end{equation}
for $\mathfrak{sp}_N$. 
This is the convention usually employed to define highest weights. A highest-weight representation contains a vector satisfying
\begin{align}
F_{i,j}^{c}\cdot\zeta & =0,\quad i<j,\\
F_{i,i}^{c}\cdot\zeta & =\lambda_{i}\zeta.
\end{align}
We denote the highest weight by $\lambda\equiv(\lambda_{1},\dots,\lambda_{n})$. The finite-dimensionality conditions are
\begin{equation}
\lambda_{i}-\lambda_{i+1}\in\mathbb{Z}_{\geq0},
\end{equation}
for $i=1,\dots,n-1$, together with
\begin{equation}
\begin{split}
2\lambda_{n} & \in\mathbb{Z}_{\geq0},\quad\text{for }\mathfrak{so}_{2n+1},\\
\lambda_{n} & \in\mathbb{Z}_{\geq0},\quad\text{for }\mathfrak{sp}_{2n},\\
\lambda_{n-1}+\lambda_{n} & \in\mathbb{Z}_{\geq0},\quad\text{for }\mathfrak{so}_{2n}.
\end{split}
\end{equation}
To compare with Molev's formulas \cite{MolevBook,Molev:1997wp,Molev:GT}, we also record the conversion to his indexing convention. For even $N$, the indices are relabelled as
\begin{equation}
(1,2,\dots,N)\to(-n,-n+1,\dots,-1,1,\dots,n-1,n),
\end{equation}
and for odd $N$ as
\begin{equation}
(1,2,\dots,N)\to(-n,-n+1,\dots,-1,0,1,\dots,n-1,n).
\end{equation}
Thus, our generators are related to those in Molev's convention by
\begin{equation}
\left(\begin{array}{ccc}
F_{1,1}^{c} & \dots & F_{1,N}^{c}\\
\vdots & \ddots & \vdots\\
F_{N,1}^{c} & \dots & F_{N,N}^{c}
\end{array}\right)=\left(\begin{array}{ccc}
F_{-n,-n}^{\mathrm{Mol}} & \dots & F_{-n,n}^{\mathrm{Mol}}\\
\vdots & \ddots & \vdots\\
F_{n,-n}^{\mathrm{Mol}} & \dots & F_{n,n}^{\mathrm{Mol}}
\end{array}\right).
\end{equation}
The corresponding highest weights are related by
\begin{equation}
(\lambda_{1}^{\mathrm{Mol}},\dots,\lambda_{n}^{\mathrm{Mol}})=(-\lambda_{n},\dots,-\lambda_{1}).
\end{equation}
The Gelfand--Tsetlin basis is particularly useful here because the relevant Sklyanin determinants and quasi-determinants are diagonal in this basis. We summarize its main properties following \cite{Molev:GT}. The basis vectors are labeled by GT patterns, specified by a collection of weights $\Lambda=\left\{ \lambda_{i,j}\right\} _{i,j}$.

For $\mathfrak{so}_{N}$, the indices range over $2\leq i\leq N$ and $1\leq j\leq\left\lfloor i/2\right\rfloor $, and the weights obey the interlacing conditions
\begin{equation}
\begin{split}
\lambda_{i,j} & \geq\lambda_{i-1,j}\geq\lambda_{i,j+1},\\
\lambda_{2i+2,i} & \geq\lambda_{2i+1,i}\geq|\lambda_{2i+2,i+1}|,\\
\lambda_{2i+1,i} & \geq\lambda_{2i,i}\geq-\lambda_{2i+1,i},
\end{split}
\end{equation}
with $\lambda_{N,j}\equiv\lambda_{j}$ and the differences between the parameters are integers
\footnote{In the symplectic case every parameter is integer and in the orthogonal case they are integers or half integers.}.
For example, an $\mathfrak{so}_{6}$ pattern has the form
\begin{equation}
\Lambda=\begin{array}{ccccc}
\lambda_{6,1} &  & \lambda_{6,2} &  & \lambda_{6,3}\\
 & \lambda_{5,1} &  & \lambda_{5,2}\\
 &  & \lambda_{4,1} &  & \lambda_{4,2}\\
 &  &  & \lambda_{3,1}\\
 &  &  &  & \lambda_{2,1}
\end{array}.
\end{equation}
The basis vector labelled by $\Lambda$ belongs to an irreducible representation of the subalgebra $\mathfrak{so}_{k}$ with highest weight $(\lambda_{k,1},\lambda_{k,2},\dots,\lambda_{k,\left\lfloor k/2\right\rfloor })$.

Each row therefore identifies the representation selected at one step of the orthogonal branching chain. A central element of the corresponding subalgebra acts by a scalar determined by that row. We will use this property to evaluate all nested Sklyanin determinants on a single GT-labelled vector.

For $\mathfrak{sp}_{2n}$, the indices range over $1\leq i\leq N$ and $1\leq j\leq\left\lceil i/2\right\rceil $, and the interlacing conditions are
\begin{equation}
\begin{split}
\lambda_{i,j} & \geq\lambda_{i-1,j}\geq\lambda_{i,j+1},\\
\lambda_{2i,i}\geq & \lambda_{2i-1,i}\geq0,\\
\lambda_{2i+1,i} & \geq\lambda_{2i,i}\geq\lambda_{2i+1,i+1},
\end{split}
\end{equation}
with $\lambda_{2n,j}\equiv\lambda_{j}$. For example, an $\mathfrak{sp}_{6}$ pattern takes the form
\begin{equation}
\Lambda=\begin{array}{cccccc}
\lambda_{6,1} &  & \lambda_{6,2} &  & \lambda_{6,3}\\
 & \lambda_{5,1} &  & \lambda_{5,2} &  & \lambda_{5,3}\\
 &  & \lambda_{4,1} &  & \lambda_{4,2}\\
 &  &  & \lambda_{3,1} &  & \lambda_{3,2}\\
 &  &  &  & \lambda_{2,1}\\
 &  &  &  &  & \lambda_{1,1}
\end{array}.
\end{equation}
We also use the following notation for the entries of symplectic GT patterns:
\begin{equation}
\tilde{\lambda}_{i,j}\equiv\lambda_{2i,j},\quad\mu_{i,j}\equiv\lambda_{2i-1,j}.\label{eq:lamt}
\end{equation}
In this notation, the $\mathfrak{sp}_{6}$ pattern becomes
\begin{equation}
\Lambda=\begin{array}{cccccc}
\tilde{\lambda}_{3,1} &  & \tilde{\lambda}_{3,2} &  & \tilde{\lambda}_{3,3}\\
 & \mu_{3,1} &  & \mu_{3,2} &  & \mu_{3,3}\\
 &  & \tilde{\lambda}_{2,1} &  & \tilde{\lambda}_{2,2}\\
 &  &  & \mu_{2,1} &  & \mu_{2,2}\\
 &  &  &  & \tilde{\lambda}_{1,1}\\
 &  &  &  &  & \mu_{1,1}
\end{array}.
\end{equation}
The basis vector labelled by $\Lambda$ belongs to an irreducible representation of the subalgebra $\mathfrak{sp}_{2k}$ with highest weight $(\lambda_{2k,1},\lambda_{2k,2},\dots,\lambda_{2k,k})=(\tilde{\lambda}_{k,1},\tilde{\lambda}_{k,2},\dots,\tilde{\lambda}_{k,k})$.

Here the even rows specify the highest weights of the nested symplectic subalgebras, while the intervening odd rows resolve the multiplicity spaces. Consequently, the even-rank Sklyanin determinants determine only part of the spectral data. The rank-two construction in Section~\ref{sec:symplectic} supplies the dependence on the intermediate rows.

\subsection{Integrable spin chains and Bethe states}

With the boundary representation conventions in place, we turn to the bulk eigenstates whose overlaps we wish to compute. We consider $\mathfrak{gl}_{N}$-symmetric spin chains in the Yangian framework.

The algebraic description starts from the Yang $R$-matrix
\begin{equation}
R(u)=1+\frac{1}{u}P,
\end{equation}
where $R(u)$ acts on $\mathbb{C}^{N}\otimes\mathbb{C}^{N}$ and $P$ permutes the two tensor factors. This $R$-matrix defines the Yangian $Y(N)$ through the RTT relation
\begin{equation}
R_{1,2}(u-v)T_{1}(u)T_{2}(v)=T_{2}(v)T_{1}(v)R_{1,2}(u-v),\label{eq:RTT}
\end{equation}
where
\begin{equation}
T(u)=\sum_{i,j=1}^{N}e_{i,j}\otimes T_{i,j}(u).
\end{equation}
Here $e_{i,j}$ are the $N\times N$ matrix units, and $T_{i,j}(u)$ are the generating series of $Y(N)$. The RTT relation (\ref{eq:RTT}) implies that the transfer matrix
\begin{equation}
\mathcal{T}(u)=\sum_{i=1}^{N}T_{i,i}(u)
\end{equation}
generates a commuting family:
\begin{equation}
\left[\mathcal{T}(u),\mathcal{T}(v)\right]=0.
\end{equation}
Let $L$ be the representation space of a highest-weight representation of $Y(N)$. Its highest-weight vector $|0\rangle\in L$ satisfies
\begin{align}
T_{i,j}(u)|0\rangle & =0,\quad i<j,\\
T_{i,i}(u)|0\rangle & =\lambda_{i}(u)|0\rangle.
\end{align}
The transfer-matrix eigenstates are parametrized by Bethe roots $u_{j}^{s}$, with $s=1,\dots,N-1$ and $j=1,\dots,n_{s}$, where $n_{s}$ is the number of roots of type $s$. Introduce the sets $\bar{u}^{s}=\left\{ u_{j}^{s}\right\} _{j=1}^{n_{s}}$ and $\bar{u}=\left\{ \bar{u}^{s}\right\} _{s=1}^{N-1}$. The roots satisfy the Bethe equations \cite{OGIEVETSKY1986360}
\begin{equation}
\frac{\lambda_{s}(iu^{s}_{k}-s/2)}{\lambda_{s+1}(iu^{s}_{k}-s/2)}=-
\prod^{n_{s}}_{j=1}\frac{u^{s}_{k}-u^{s}_{j}-i}{u^{s}_{k}-u^{s}_{j}+i}
\prod^{n_{s-1}}_{j=1}\frac{u^{s}_{k}-u^{s-1}_{j}+i/2}{u^{s}_{k}-u^{s-1}_{j}-i/2}
\prod^{n_{s+1}}_{j=1}\frac{u^{s}_{k}-u^{s+1}_{j}+i/2}{u^{s}_{k}-u^{s+1}_{j}-i/2},
\end{equation}
for all $s=1,\dots,N-1$ and $k=1,\dots,n_{s}$. The corresponding Bethe vectors $|\bar{u}\rangle$ satisfy
\begin{equation}
\mathcal{T}(v)|\bar{u}\rangle=\tau(v|\bar{u})|\bar{u}\rangle,
\end{equation}
with eigenvalue
\begin{equation}
\tau(v|\bar{u})=\sum_{s=1}^{N}\lambda_{s}(v)\prod^{n_{s}}_{j=1}\frac{v-iu^{s}_{j}+s/2-1}{v-iu^{s}_{j}+s/2}\prod^{n_{s-1}}_{j=1}\frac{v-iu^{s-1}_{j}+(s+1)/2}{v-iu^{s-1}_{j}+(s-1)/2},
\end{equation}
where $n_0=n_N=0$.
The highest-weight functions of the monodromy matrix encode the quantum space and its inhomogeneities. They are distinct from the constant Lie-algebra weights of the boundary representation: the former enter the Bethe equations, while the latter determine the F-functions.

We also need the inverse-transposed monodromy matrix, defined by
\begin{equation}
\widehat{T}_{j,i}(u)T_{j,k}(u)=\delta_{i,k},
\end{equation}
or, in matrix notation,
\begin{equation}
\widehat{T}^{t}(u)T(u)=1,
\end{equation}
where $^{t}$ denotes transposition in the auxiliary space.

\subsection{Integrable matrix product states and overlaps}

We define integrable MPSs through the KT relation \cite{Gombor:2024iix,Gombor:2025wvu}. Consider $d^{2}$ covectors $\langle\psi_{\alpha,\beta}|$ in the quantum space. They define an integrable boundary state if
\begin{equation}
\sum_{j=1}^{N}\sum_{\beta=1}^{d}K_{i,j}^{\alpha,\beta}(z)\langle\psi_{\beta,\gamma}|T_{j,k}(z)=\sum_{j=1}^{N}\sum_{\beta=1}^{d}\langle\psi_{\alpha,\beta}|\widehat{T}_{i,j}(-z)K_{j,k}^{\beta,\gamma}(z),
\end{equation}
for all $i,k=1,\dots,N$ and $\alpha,\gamma=1,\dots,d$.\footnote{This is the crossed KT relation. An uncrossed version also exists, but we restrict attention to the crossed case.} Here $K_{i,j}^{\alpha,\beta}(z)$ are scalar functions of the spectral parameter. We organize these quantities into matrices acting on the boundary space $\mathcal{V}_{B}\cong\mathbb{C}^{d}$. Let $e_{\alpha,\beta}^{B}$, with $\alpha,\beta=1,\dots,d$, be the matrix units in this space, and define
\begin{equation}
\langle\Psi|=\sum_{\alpha,\beta=1}^{d}\langle\psi_{\alpha,\beta}|\otimes e_{\alpha,\beta}^{B},\qquad\mathbf{K}_{i,j}(z)=\sum_{\alpha,\beta=1}^{d}K_{i,j}^{\alpha,\beta}(z)e_{\alpha,\beta}^{B}.
\end{equation}
Thus $\mathbf{K}_{i,j}(z)\in\mathrm{End}(\mathbb{C}^{d})$ acts on the boundary space for each $i,j=1,\dots,N$. In this notation, the KT relation reads
\begin{equation}
\sum_{j=1}^{N}\mathbf{K}_{i,j}(z)\langle\Psi|T_{j,k}(z)=\sum_{j=1}^{N}\langle\Psi|\widehat{T}_{i,j}(-z)\mathbf{K}_{j,k}(z).
\end{equation}
The KT relation intertwines the ordinary and crossed monodromies through the boundary matrix and can be imposed before specifying the local spin representations. It also provides the algebraic basis for the matrix product construction. In the products below, multiplication is understood in the shared auxiliary or boundary space, while the quantum factors act on distinct sites or blocks of sites.

To see how this construction works, recall the coproduct property of the monodromy matrix: if $T^{(1)}(z),T^{(2)}(z),\dots,T^{(J)}(z)$ satisfy the RTT relation with the same $R$-matrix, then so does the matrix product operator (MPO)
\begin{equation}
T_{0}(z)=T_{0}^{(1)}(z)\otimes T_{0}^{(2)}(z)\otimes\dots\otimes T_{0}^{(J)}(z).
\end{equation}
The KT relation has an analogous property. If $\langle\Psi^{(1)}|,\langle\Psi^{(2)}|,\dots,\langle\Psi^{(J)}|$ satisfy it with the same $K$-matrix, so that
\begin{equation}
\sum_{j=1}^{N}\mathbf{K}_{i,j}(z)\langle\Psi^{(a)}|T_{j,k}^{(a)}(z)=\sum_{j=1}^{N}\langle\Psi^{(a)}|\widehat{T}_{i,j}^{(a)}(-z)\mathbf{K}_{j,k}(z),
\end{equation}
for all $a=1,\dots,J$, then the matrix product state
\begin{equation}
\langle\Psi|=\langle\Psi^{(1)}|\otimes\langle\Psi^{(2)}|\otimes\dots\otimes\langle\Psi^{(J)}|
\end{equation}
also satisfies the KT relation. This construction gives an untraced MPS, which is still a matrix in the boundary space. Taking its trace, in analogy with the transfer matrix, gives
\begin{equation}
\langle\mathrm{MPS}|=\sum_{\alpha=1}^{d}\langle\psi_{\alpha,\alpha}|,
\end{equation}
which is a covector in the quantum space.

The compatibility of the KT relation also leads to the reflection equation for the $K$-matrix:
\begin{equation}
R_{1,2}(v-u)\mathbf{K}_{1}(u)\bar{R}_{1,2}(-u-v)\mathbf{K}_{2}(v)=\mathbf{K}_{2}(v)\bar{R}_{1,2}(-u-v)\mathbf{K}_{1}(u)R_{1,2}(v-u),
\end{equation}
where
\begin{equation}
\bar{R}_{1,2}(u)=R_{1,2}^{T_{1}}(-u),\qquad\mathbf{K}(u)=\sum_{i,j=1}^{N}e_{i,j}\otimes\mathbf{K}_{i,j}(u).
\end{equation}
Together with a symmetry relation, this equation defines the twisted Yangians $Y^{\pm}(N)$, and the $K$-matrix realizes a representation of the corresponding algebra. We focus on evaluation representations, a broad class obtained from finite-dimensional irreducible representations of the orthogonal and symplectic Lie algebras. Their construction is reviewed below.

Another important consequence of the KT relation is the pair structure of Bethe roots: the overlap
\begin{equation}
\langle\mathrm{MPS}|\bar{u}\rangle
\end{equation}
can be nonzero only if the Bethe roots occur in pairs \cite{Gombor:2024iix,Gombor:2025wvu}. More precisely, for each type $s$,
\begin{equation}
u_{j}^{s}=-u_{n_{s}+1-j}^{s}.
\end{equation}
Throughout this paper we assume that every $n_{s}$ is even.

The F-functions entering (\ref{eq:overlap}) are obtained from the G-operators. Their definitions differ in the $\mathfrak{so}_{N}$ and $\mathfrak{sp}_{2n}$ cases and will be given separately below. The F-operators are defined by
\begin{equation}
\mathbf{F}^{(k)}=\left(\mathbf{G}^{(k)}\right)^{-1}\mathbf{G}^{(k+1)}.
\end{equation}
These operators act in the finite-dimensional boundary space, with inverses understood at generic spectral parameter, or equivalently as meromorphic operator-valued functions. They form a commuting family, so in a common eigenbasis their ratios reduce to scalar ratios. A representation-theoretic description of this basis can therefore replace explicit matrix inversion.

Writing the simultaneous diagonalization explicitly, we have
\begin{equation}
\mathbf{G}^{(k)}=\mathbf{A}\mathrm{diag}\left(\mathcal{G}_{1}^{(k)},\dots,\mathcal{G}_{d}^{(k)}\right)\mathbf{A}^{-1},\quad\mathbf{F}^{(k)}=\mathbf{A}\mathrm{diag}\left(\mathcal{F}_{1}^{(k)},\dots,\mathcal{F}_{d}^{(k)}\right)\mathbf{A}^{-1}.
\end{equation}
The functions entering the overlap formula are obtained by the reparametrization
\begin{equation}
\tilde{\mathcal{F}}_{j}^{(k)}(u)=\mathcal{F}_{j}^{(k)}(iu-k/2)\sqrt{\frac{u^{2}}{u^{2}+1/4}}.
\end{equation}
\subsection{Properties of the twisted Yangians}

We now introduce the Sklyanin minors and determinants used to determine these eigenvalues. Let $u_{1},\dots,u_{k}$ be independent variables and define
\begin{equation}
R(u_{1},\dots,u_{k})=\left(R_{k-1,k}\right)\left(R_{k-2,k}R_{k-2,k-1}\right)\dots\left(R_{1,k}\dots R_{1,2}\right),
\end{equation}
where
\begin{equation}
R_{i,j}\equiv R_{i,j}(u_{j}-u_{i}).
\end{equation}
Similarly, set
\begin{equation}
\mathbf{K}_{i}\equiv\mathbf{K}_{i}(u_{i}),\quad\bar{R}_{i,j}\equiv\bar{R}_{i,j}(-u_{i}-u_{j}).
\end{equation}
The following fused product keeps track of the ordering of the boundary matrices and crossed $R$-matrices. This ordering is essential because their entries act on the same boundary space and need not commute. Antisymmetrizing the auxiliary factors will produce the minors whose maximal component is the Sklyanin determinant.

For any permutation $(p_{1},\dots,p_{k})$ of $1,\dots,k$, define
\begin{equation}
\left\langle \mathbf{K}_{p_{1}},\dots,\mathbf{K}_{p_{k}}\right\rangle =\mathbf{K}_{p_{1}}\left(\bar{R}_{p_{1},p_{2}}\dots\bar{R}_{p_{1},p_{k}}\right)\mathbf{K}_{p_{2}}\left(\bar{R}_{p_{2},p_{3}}\dots\bar{R}_{p_{2},p_{k}}\right)\dots\mathbf{K}_{p_{k-1}}\left(\bar{R}_{p_{k-1},p_{k}}\right)\mathbf{K}_{p_{k}}.
\end{equation}
The boundary Yang--Baxter equation implies
\begin{equation}
R(u_{1},\dots,u_{k})\left\langle \mathbf{K}_{1},\dots,\mathbf{K}_{k}\right\rangle =\left\langle \mathbf{K}_{k},\dots,\mathbf{K}_{1}\right\rangle R(u_{1},\dots,u_{k}).
\end{equation}
We now specialize the spectral parameters to
\begin{equation}
u_{i}=u-i+1.
\end{equation}
Under this specialization, $R(u_{1},\dots,u_{k})$ becomes the antisymmetrizer on $(\mathbb{C}^{N})^{\otimes k}$:
\begin{equation}
A_{k}=\sum_{\sigma\in\mathcal{S}_{k}}\mathrm{sgn}\sigma\cdot P_{\sigma}.
\end{equation}
Here $P_{\sigma}$ denotes the natural action of the permutation $\sigma\in\mathcal{S}_{k}$ on $(\mathbb{C}^{N})^{\otimes k}$. Consequently,
\begin{equation}
A_{k}\left\langle \mathbf{K}_{1},\dots,\mathbf{K}_{k}\right\rangle =\left\langle \mathbf{K}_{k},\dots,\mathbf{K}_{1}\right\rangle A_{k}.
\end{equation}
We introduce component notation by writing
\begin{equation}
\sum_{a_{j},b_{j}}e_{a_{1},b_{1}}\otimes\dots\otimes e_{a_{k},b_{k}}\otimes\mathbf{K}_{a_{1},\dots,a_{k}}^{b_{1},\dots,b_{k}}(u)=A_{k}\left\langle \mathbf{K}_{1},\dots,\mathbf{K}_{k}\right\rangle =\left\langle \mathbf{K}_{k},\dots,\mathbf{K}_{1}\right\rangle A_{k}.
\end{equation}
The coefficients $\mathbf{K}_{a_{1},\dots,a_{k}}^{b_{1},\dots,b_{k}}(u)$ are the Sklyanin minors \cite{MolevBook}. They are antisymmetric under permutations of either the upper or the lower indices:
\begin{equation}
\mathbf{K}_{a_{1},\dots,a_{k}}^{b_{\sigma(1)},\dots,b_{\sigma(k)}}(u)=\mathrm{sgn}\sigma\cdot\mathbf{K}_{a_{1},\dots,a_{k}}^{b_{1},\dots,b_{k}}(u),\quad\mathbf{K}_{a_{\sigma(1)},\dots,a_{\sigma(k)}}^{b_{1},\dots,b_{k}}(u)=\mathrm{sgn}\sigma\cdot\mathbf{K}_{a_{1},\dots,a_{k}}^{b_{1},\dots,b_{k}}(u),
\end{equation}
for all $\sigma\in\mathcal{S}_{k}$.

At maximal degree, the antisymmetric auxiliary space is one-dimensional, so the fused operator has only one independent auxiliary component.

This component defines the Sklyanin determinant:
\begin{equation}
\mathrm{sdet}\mathbf{K}(u):=\mathbf{K}_{1,\dots,N}^{1,\dots,N}(u).
\end{equation}
The coefficients of the Sklyanin determinant are central in the twisted Yangian. They therefore act as scalars on an irreducible representation, and their eigenvalues can be found by evaluating them on a highest-weight vector.

On an irreducible highest-weight representation, the Sklyanin determinant is given by \cite[Proposition 4.2.4]{MolevBook}
\begin{equation}
\mathrm{sdet}\mathbf{K}(u)=\alpha_{N}(u)\prod_{j=1}^{N}\tilde{\mu}_{j}(u-j+1),\label{eq:sdetmu}
\end{equation}
where
\begin{equation}
\alpha_{N}(u)=\begin{cases}
1, & \text{for }\mathfrak{so}_{N},\\
\frac{2u+1}{2u-N+1}, & \text{for }\mathfrak{sp}_{N}.
\end{cases}
\end{equation}
The highest-weight conventions are specified in Appendix~\ref{app:conventions}.

We focus on $K$-matrices associated with the evaluation homomorphisms
\begin{equation}
\mathbf{K}_{i,j}(u)=\begin{cases}
\delta_{i,j}+\frac{1}{u+1/2}F_{i,j}, & \text{for }\mathfrak{so}_{N},\\
(-1)^i\delta_{\bar{i},j}+(-1)^i \frac{1}{u-1/2}F_{\bar{i},j}, & \text{for }\mathfrak{sp}_{N}.
\end{cases}
\end{equation}
For these representations, the Sklyanin determinant is
\begin{equation}
\mathrm{sdet}\mathbf{K}(u)=\alpha_{N}(u)\prod_{j=1}^{n}\frac{u-j+1\mp1/2+\lambda_{j}}{u-j+1\mp1/2}\prod_{j=1}^{n}\frac{u-N+j\pm1/2-\lambda_{j}}{u-N+j\pm1/2},\label{eq:evalSdet}
\end{equation}
where we used $\lambda_{N+1-i}=-\lambda_{i}$.

We denote the $K$-matrix of the subalgebra $Y^{+}(k)$ by $\mathbf{K}^{[k]}(u)$, with
\begin{equation}
\mathbf{K}_{i,j}^{[k]}(u)=\mathbf{K}_{i,j}(u),\quad i,j=1,\dots,k.
\end{equation}
The Sklyanin determinants of the nested subalgebras $Y^{\pm}(k)$ can also be written as Sklyanin minors. We use the notation
\begin{equation}
\mathbf{D}^{(k)}(u):=\mathrm{sdet}\mathbf{K}^{[k]}(u)=\mathbf{K}_{1,\dots,k}^{1,\dots,k}(u).
\end{equation}
\section{Overlaps for $\mathfrak{so}_{N}$-symmetric MPSs}\label{sec:orthogonal}

This section derives the overlap functions for the orthogonal family of boundary representations. Starting from the quasi-determinant definition of the G-operators, we use a Sklyanin-determinant factorization to express their spectra through central elements of nested subalgebras. This avoids evaluating operator-valued inverses explicitly and makes the representation dependence accessible through highest weights. For evaluation representations, the GT branching labels then give the overlap functions in closed form. We conclude by explaining how the same determinant ratios apply to more general $K$-matrices when the highest-weight data of their nested restrictions are available.

\subsection{Overlaps with quasi-determinants}

In the orthogonal case, the G-operators are defined by the quasi-determinants
\begin{multline}
\mathbf{G}^{(k)}=\left|\begin{array}{cccc}
\mathbf{K}_{1,1} & \mathbf{K}_{1,2} & \dots & \mathbf{K}_{1,k}\\
\mathbf{K}_{2,1} & \mathbf{K}_{2,2} & \dots & \mathbf{K}_{2,k}\\
\vdots & \vdots & \ddots & \vdots\\
\mathbf{K}_{k,1} & \mathbf{K}_{k,2} & \dots & \boxed{\mathbf{K}_{k,k}}
\end{array}\right| :=\\
\mathbf{K}_{k,k}-\left(\begin{array}{ccc}
\mathbf{K}_{k,1} & \dots & \mathbf{K}_{k,k-1}\end{array}\right)\left(\begin{array}{ccc}
\mathbf{K}_{1,1} & \dots & \mathbf{K}_{1,k-1}\\
\vdots & \ddots & \vdots\\
\mathbf{K}_{k-1,1} & \dots & \mathbf{K}_{k-1,k-1}
\end{array}\right)^{-1}\left(\begin{array}{c}
\mathbf{K}_{1,k}\\
\vdots\\
\mathbf{K}_{k-1,k}
\end{array}\right).
\end{multline}
Quasi-determinants have been extensively studied in mathematics, particularly in noncommutative algebra \cite{Gelfand1991DeterminantsOM,Gelfand:1994hg,molev2003quasi,GELFAND200556}. In the following, we summarize the relations between quasi-determinants and Sklyanin determinants of twisted Yangians.

\subsection{Overlaps with Sklyanin determinants}

The quasi-determinant above is the noncommutative Schur complement of the leading auxiliary block. Rather than invert this block explicitly, we can recover the same operator from two consecutive Sklyanin determinants, whose eigenvalues are fixed by highest weights.

The quasi-determinants and Sklyanin determinants obey the factorization identity
\begin{equation}
\mathbf{D}^{(k)}(u)=\mathbf{D}^{(k-1)}(u)\mathbf{G}^{(k)}(u-k+1).\label{eq:DDG}
\end{equation}
This identity follows from the factorization theorem in \cite[Theorem 2.12.1]{MolevBook}. For completeness, we give a proof in Appendix~\ref{app:factorization}. Solving for the G-operators gives
\begin{equation}
\mathbf{G}^{(k)}(u)=\left[\mathbf{D}^{(k-1)}(u+k-1)\right]^{-1}\mathbf{D}^{(k)}(u+k-1).
\end{equation}
To pass to eigenvalues, choose a common eigenvector of the nested determinant family. Although a nested determinant need not be central in the full algebra, it is central in its own subalgebra; the nested central elements commute with one another. Evaluating their ratio on a common eigenvector gives
\begin{equation}
\mathcal{G}_{j}^{(k)}(u)=\frac{\mathcal{D}_{j}^{(k)}(u+k-1)}{\mathcal{D}_{j}^{(k-1)}(u+k-1)},\label{eq:Geig}
\end{equation}
and
\begin{equation}
\mathcal{F}_{j}^{(k)}(u)=\frac{\mathcal{D}_{j}^{(k-1)}(u+k-1)\mathcal{D}_{j}^{(k+1)}(u+k)}{\mathcal{D}_{j}^{(k)}(u+k-1)\mathcal{D}_{j}^{(k)}(u+k)}.\label{eq:Feig}
\end{equation}
\subsection{Application to evaluation representations}

We now apply these relations to the evaluation $K$-matrix
\footnote{Here we concentrate on $N>2$ cases. For $N=2$ the evaluation $K$-matrices are scalars in the boundary space for which the overlaps have been already derived in \cite{Gombor:2021uxz}.}
\begin{equation}
\mathbf{K}_{i,j}(u)=\delta_{i,j}+\frac{1}{u+1/2}F_{i,j},
\end{equation}
where $F_{i,j}$ are the $\mathfrak{so}_{N}$ generators defined in Section~\ref{sec:Representations-of-the}. For the representation with highest weight $(\lambda_{1},\lambda_{2},\dots,\lambda_{n})$, the Sklyanin determinant is given by (\ref{eq:evalSdet}):
\begin{equation}
\mathrm{sdet}\mathbf{K}(u)=\prod_{j=1}^{n}\frac{u-j+1/2+\lambda_{j}}{u-j+1/2}\frac{-u+N-j-1/2+\lambda_{j}}{-u+N-j-1/2}.
\end{equation}
In the GT basis, the Sklyanin determinants of the subalgebras $Y^{+}(k)$ have eigenvalues
\begin{equation}
\mathcal{D}_{\Lambda}^{(k)}(u)=\prod_{j=1}^{\left\lfloor k/2\right\rfloor }\frac{u-j+1/2+\lambda_{k,j}}{u-j+1/2}\frac{-u+k-j-1/2+\lambda_{k,j}}{-u+k-j-1/2}.
\end{equation}
For a fixed GT pattern, the row at level $k$ supplies the highest weight of the rank-$k$ determinant. Thus (\ref{eq:Geig}) compares two adjacent rows, while (\ref{eq:Feig}) involves three consecutive rows. Since the row length changes at every other step, we separate even and odd levels. Simplifying the weight-independent factors gives
\begin{equation}
\mathcal{G}_{\Lambda}^{(k)}(u)=\begin{cases}
\frac{1}{(u+l-1/2)(-u-1/2)}\frac{\prod_{j=1}^{l}(u+k-j-1/2+\lambda_{k,j})(-u-j+1/2+\lambda_{k,j})}{\prod_{j=1}^{l-1}(u+k-j-1/2+\lambda_{k-1,j})(-u-j-1/2+\lambda_{k-1,j})}, & k=2l,\\
\frac{u+l+1/2}{u+1/2}\frac{\prod_{j=1}^{l}(u+k-j-1/2+\lambda_{k,j})(-u-j+1/2+\lambda_{k,j})}{\prod_{j=1}^{l}(u+k-j-1/2+\lambda_{k-1,j})(-u-j-1/2+\lambda_{k-1,j})}, & k=2l+1,
\end{cases}
\end{equation}
and
\begin{multline}
\mathcal{F}_{\Lambda}^{(k)}(u)=
a_k(u) \frac{\prod_{j=1}^{\left\lfloor \frac{k-1}{2}\right\rfloor}(u+k-j-1/2+\lambda_{k-1,j})(u+j+1/2-\lambda_{k-1,j})}{\prod_{j=1}^{\left\lfloor k/2\right\rfloor}(u+k-j-1/2+\lambda_{k,j})(u+j-1/2-\lambda_{k,j})} \times \\
\frac{\prod_{j=1}^{\left\lfloor \frac{k+1}{2}\right\rfloor}(u+k-j+1/2+\lambda_{k+1,j})(u+j-1/2-\lambda_{k+1,j})}{\prod_{j=1}^{\left\lfloor k/2\right\rfloor}(u+k-j+1/2+\lambda_{k,j})(u+j+1/2-\lambda_{k,j})}, 
\end{multline}
where
\begin{equation}
a_k(u)=
\begin{cases}
\frac{1}{(u+k/2)^{2}}, & k=2l-1,\\
(u+k/2)^{2}-1/4, & k=2l.
\end{cases}
\end{equation}
To recover the functions used in the overlap formula, we now apply the spectral reparametrization and scalar normalization defined in Section~\ref{sec:preliminaries}. The paired linear factors combine into expressions quadratic in the rapidity, yielding
\begingroup\fontsize{9.5}{11.5}\selectfont
\begin{equation}
\tilde{\mathcal{F}}_{\Lambda}^{(k)}(u)=\Upsilon_{k}(u)
\frac{\prod_{j=1}^{\left\lfloor \frac{k-1}{2}\right\rfloor }(u^{2}+(\lambda_{k-1,j}+k/2-j-1/2)^{2})}{\prod_{j=1}^{\left\lfloor k/2\right\rfloor }(u^{2}+(\lambda_{k,j}+k/2-j-1/2)^{2})}
\frac{\prod_{j=1}^{\left\lfloor \frac{k+1}{2}\right\rfloor }(u^{2}+(\lambda_{k+1,j}+k/2-j+1/2)^{2})}{\prod_{j=1}^{\left\lfloor k/2\right\rfloor }(u^{2}+(\lambda_{k,j}+k/2-j+1/2)^{2})},
\end{equation}
\endgroup
where
\begin{equation}
\Upsilon_{k}(u)=\begin{cases}
\frac{1}{\sqrt{u^{2}(u^{2}+1/4)}}, & k=2l-1,\\
\sqrt{u^{2}(u^{2}+1/4)}, & k=2l.
\end{cases}
\end{equation}
Introducing the shifted weights
\begin{equation}
\gamma_{k,j}=\lambda_{k,j}+\frac{k}{2}-j-\frac{1}{2},
\end{equation}
we can write the result in the compact form
\begin{equation}
\tilde{\mathcal{F}}_{\Lambda}^{(k)}(u)=\Upsilon_{k}(u)\frac{\prod_{j=1}^{\left\lfloor \frac{k-1}{2}\right\rfloor }(u^{2}+(\gamma_{k-1,j}+1/2)^{2})}{\prod_{j=1}^{\left\lfloor k/2\right\rfloor }(u^{2}+\gamma_{k,j}^{2})}\frac{\prod_{j=1}^{\left\lfloor \frac{k+1}{2}\right\rfloor }(u^{2}+(\gamma_{k+1,j}+1/2)^{2})}{\prod_{j=1}^{\left\lfloor k/2\right\rfloor }(u^{2}+(\gamma_{k,j}+1)^{2})}.
\end{equation}
\subsection{Overlaps for general $K$-matrices}

The preceding derivation relies on evaluation representations only through the explicit determinant eigenvalues. The factorization itself holds in the boundary algebra, so the same method applies once the branching data of a more general representation are known.

The expression (\ref{eq:Feig}) for the F-eigenvalues in terms of nested Sklyanin determinants also applies beyond evaluation representations. On an irreducible representation of $Y^{+}(N)$, the Sklyanin determinant is determined by the highest weights through (\ref{eq:sdetmu}). The functions $\tilde{\mu}_{r}(u)$ are defined in Appendix~\ref{app:conventions}.

The restriction $\mathbf{K}^{[N-1]}(u)$ is generally reducible. Denote its irreducible blocks by $V_{N-1}^{(l)}$ and the corresponding highest-weight functions by $\tilde{\mu}_{i}^{(N-1,l)}(u)$, where $l$ labels the blocks. More generally, denote the irreducible blocks of $\mathbf{K}^{[k]}(u)$ by $V_{k}^{(l)}$, with highest weights $\mu_{i}^{(k,l)}(u)$. On each block, the Sklyanin determinant acts as a scalar:
\begin{equation}
\mathbf{D}^{(k)}(u)\Biggr|_{V_{k}^{(l)}}=\mathcal{D}_{l}^{(k)}(u)\mathbf{1}.
\end{equation}
Equation (\ref{eq:sdetmu}) gives
\begin{equation}
\mathcal{D}_{l}^{(k)}(u)=\prod_{r=1}^{k}\tilde{\mu}_{r}^{(k,l)}(u-r+1).
\end{equation}
This description is to be understood for restrictions admitting the stated block decomposition and a compatible common eigenbasis. If a block occurs with multiplicity, the corresponding multiplicity labels must also be retained when summing over boundary-space states. The determinant eigenvalues alone need not distinguish those states.

In a basis adapted to these nested blocks, all $\mathbf{D}^{(k)}$ are diagonal. Each vector $|\ell\rangle$ is associated with a sequence $\ell=(l_{1},\dots,l_{N})$ specifying its block at every level:
\begin{equation}
|\ell\rangle\in V_{k}^{(l_{k})}.
\end{equation}
The corresponding Sklyanin-determinant eigenvalue is
\begin{equation}
\mathbf{D}^{(k)}(u)|\ell\rangle=\mathcal{D}_{l_{k}}^{(k)}(u)|\ell\rangle.
\end{equation}
The F-operator eigenvalues then follow as
\begin{equation}
\mathbf{F}^{(k)}(u)|\ell\rangle=\mathcal{F}_{\ell}^{(k)}(u)|\ell\rangle,
\end{equation}
where
\begingroup\fontsize{8.6}{10.6}\selectfont
\begin{equation}
\mathcal{F}_{\ell}^{(k)}(u)=\frac{\mathcal{D}_{l_{k-1}}^{(k-1)}(u+k-1)\mathcal{D}_{l_{k+1}}^{(k+1)}(u+k)}{\mathcal{D}_{l_k}^{(k)}(u+k-1)\mathcal{D}_{l_k}^{(k)}(u+k)}=
\frac{\prod_{r=1}^{k-1}\tilde{\mu}_{r}^{(k-1,l_{k-1})}(u+k-r)\prod_{r=1}^{k+1}\tilde{\mu}_{r}^{(k+1,l_{k+1})}(u+k-r+1)}{\prod_{r=1}^{k}\tilde{\mu}_{r}^{(k,l_k)}(u+k-r)\tilde{\mu}_{r}^{(k,l_k)}(u+k-r+1)}.
\end{equation}
\endgroup
The overlap formula therefore reads
\begin{equation}
\frac{\langle\mathrm{MPS}|\bar{u}\rangle}{\sqrt{\langle\bar{u}|\bar{u}\rangle}}=\sum_{\ell}\beta_{\ell}\prod_{k=1}^{N-1}\prod_{j=1}^{n_{k}/2}\tilde{\mathcal{F}}_{\ell}^{(k)}(u_{j}^{k})\times\sqrt{\frac{\det G^{+}}{\det G^{-}}},
\end{equation}
with
\begin{equation}
\tilde{\mathcal{F}}_{\ell}^{(k)}(u)=\sqrt{\frac{u^{2}}{u^{2}+1/4}}\frac{\prod_{r=1}^{k-1}\tilde{\mu}_{r}^{(k-1,l_{k-1})}(iu+k/2-r)\prod_{r=1}^{k+1}\tilde{\mu}_{r}^{(k+1,l_{k+1})}(iu+k/2-r+1)}{\prod_{r=1}^{k}\tilde{\mu}_{r}^{(k,l_k)}(iu+k/2-r)\tilde{\mu}_{r}^{(k,l_k)}(iu+k/2-r+1)}.
\end{equation}
\section{Overlaps for $\mathfrak{sp}_{2n}$-symmetric MPSs}\label{sec:symplectic}

This section develops the overlap functions for symplectic evaluation representations. The undeformed odd-level operators $\mathbf{G}^{(2k-1)}$ and $\mathbf{D}^{(2k-1)}$ are not invertible, so the one-step construction of Section~\ref{sec:orthogonal} must be replaced by a two-step nesting. We first introduce rank-two boundary matrices and the deformation that defines the appropriate G-operators. We then use even-rank Sklyanin determinants to determine the product of each adjacent pair of G-eigenvalues. The remaining information is obtained by factoring the rank-two matrices on symplectic multiplicity spaces and identifying their spectral labels with the intermediate rows of GT patterns. Combining these ingredients yields a uniform product formula for the symplectic overlap functions.

\subsection{Overlaps with quasi-determinants}

We group adjacent auxiliary directions into pairs and eliminate the preceding even-dimensional block. The resulting Schur complement retains a two-dimensional auxiliary space rather than a single pivot. Following \cite{Gombor:2024iix,Gombor:2025wvu}, the F-functions are constructed from the entries of these nested matrices. First, define
\begin{equation}
\mathbf{k}_{\alpha,\beta}^{(k)}=\left|\begin{array}{ccccc}
\mathbf{K}_{1,1} & \mathbf{K}_{1,2} & \dots & \mathbf{K}_{1,2k-2} & \mathbf{K}_{1,2k-2+\beta}\\
\mathbf{K}_{2,1} & \mathbf{K}_{2,2} & \dots & \mathbf{K}_{2,2k-2} & \mathbf{K}_{2,2k-2+\beta}\\
\vdots & \vdots & \ddots & \vdots & \vdots\\
\mathbf{K}_{2k-2,1} & \mathbf{K}_{2k-2,2} & \dots & \mathbf{K}_{2k-2,2k-2} & \mathbf{K}_{2k-2,2k-2+\beta}\\
\mathbf{K}_{2k-2+\alpha,1} & \mathbf{K}_{2k-2+\alpha,2} & \dots & \mathbf{K}_{2k-2+\alpha,2k-2} & \boxed{\mathbf{K}_{2k-2+\alpha,2k-2+\beta}}
\end{array}\right|,
\end{equation}
where $\alpha,\beta=1,2$. The matrices $\mathbf{k}^{(k)}(u-k+1)$ act on a two-dimensional auxiliary space and satisfy the boundary Yang--Baxter equation; they are $Y^{-}(2)$ $K$-matrices. To construct the G-operators, we deform these matrices while preserving that equation. This requires the factorization
\begin{equation}
\mathbf{k}_{\alpha,\beta}^{(k)}(u) = \rho^{(k)}(u)
\sum_{\gamma,\delta=1}^{2}\mathbf{L}_{\gamma,\alpha}^{(k)}(u+k-1)\epsilon_{\gamma,\delta}\mathbf{L}_{\delta,\beta}^{(k)}(-u-k+1),
\end{equation}
where $\mathbf{L}_{\alpha,\beta}^{(k)}(u)$ realizes a representation of $Y(2)$ and $\rho^{(k)}(u)$ is a scalar normalization factor. The deformed $K$-matrices are
\begin{equation}
\tilde{\mathbf{k}}_{\alpha,\beta}^{(k)}(u)=\mathbf{k}_{\alpha,\beta}^{(k)}(u)-
\rho^{(k)}(u) (u+k-1/2)\mathbf{L}_{1,\alpha}^{(k)}(u+k-1)\mathbf{L}_{1,\beta}^{(k)}(-u-k+1).
\end{equation}
The deformation supplies the odd-level pivot required for the next Schur-complement step while preserving the reflection equation. Each deformed rank-two matrix then yields a pair of consecutive G-operators:

\begin{equation}
\begin{split}\mathbf{G}^{(2k-1)}(u) & =\tilde{\mathbf{k}}_{1,1}^{(k)}(u),\\
\mathbf{G}^{(2k)}(u) & =\left|\begin{array}{cc}
\tilde{\mathbf{k}}_{1,1}^{(k)}(u) & \tilde{\mathbf{k}}_{1,2}^{(k)}(u)\\
\tilde{\mathbf{k}}_{2,1}^{(k)}(u) & \boxed{\tilde{\mathbf{k}}_{2,2}^{(k)}(u)}
\end{array}\right|.
\end{split}
\end{equation}
\subsection{Overlaps with Sklyanin determinants}

Denote the Sklyanin determinants of the nested $Y^{-}(2)$ $K$-matrices by $\mathbf{d}^{(k)}$. They can be expressed through the G-operators as
\begin{equation}
\mathbf{d}^{(k)}(u)=\mathbf{G}^{(2k-1)}(u)\mathbf{G}^{(2k)}(u-1),
\end{equation}
and through the Sklyanin determinants of the even-rank subalgebras as
\begin{equation}
\mathbf{D}^{(2k)}(u)=\mathbf{D}^{(2k-2)}(u)\mathbf{d}^{(k)}(u-2k+2).\label{eq:DDd}
\end{equation}
The first relation fixes the product of two adjacent G-operators; the second expresses it through the central elements of two consecutive even-rank subalgebras. These determinant data thus determine the pair as a whole, leaving one of its factors to be found separately.

The proof parallels that of (\ref{eq:DDG}) and is given in Appendix~\ref{app:factorization}. At the level of eigenvalues, we obtain
\begin{equation}
d_{j}^{(k)}(u)=\frac{\mathcal{D}_{j}^{(2k)}(u+2k-2)}{\mathcal{D}_{j}^{(2k-2)}(u+2k-2)},\label{eq:Ym2d}
\end{equation}
and
\begin{equation}
\mathcal{G}_{j}^{(2k)}(u)=\frac{d_{j}^{(k)}(u+1)}{\mathcal{G}_{j}^{(2k-1)}(u+1)}=\frac{\mathcal{D}_{j}^{(2k)}(u+2k-1)}{\mathcal{G}_{j}^{(2k-1)}(u+1)\mathcal{D}_{j}^{(2k-2)}(u+2k-1)}.\label{eq:G2k}
\end{equation}
It therefore remains to determine the odd-level eigenvalues $\mathcal{G}_{j}^{(2k-1)}$; together with the determinant eigenvalues $\mathcal{D}_{j}^{(l)}$, they fix the even-level eigenvalues $\mathcal{G}_{j}^{(2k)}$.

For this remaining step, we will also need the expression of the nested $Y^{-}(2)$ $K$-matrices in terms of Sklyanin minors:
\begin{equation}
\mathbf{K}_{1,2,\dots,2k-2,2k-2+\alpha}^{1,2,\dots,2k-2,2k-2+\beta}(u)=\mathbf{D}^{(2k-2)}(u)\mathbf{k}_{\alpha,\beta}^{(k)}(u-2k+2).\label{eq:connKk}
\end{equation}
\subsection{Application to evaluation representations}

We apply these relations to the evaluation $K$-matrix
\begin{equation}
\mathbf{K}_{i,j}(u) = (-1)^i\delta_{\bar{i},j}+(-1)^i \frac{1}{u-1/2}F_{\bar{i},j},
\end{equation}
where $F_{i,j}$ are the $\mathfrak{sp}_{2n}$ generators defined in Section~\ref{sec:Representations-of-the}. For the representation with highest weight $(\lambda_{1},\lambda_{2},\dots,\lambda_{n})$, equation (\ref{eq:evalSdet}) gives
\begin{equation}
\mathrm{sdet}\mathbf{K}(u)=\alpha_{N}(u)\prod_{i=1}^{n}\frac{u+\lambda_{i}+3/2-i}{u+3/2-i}\frac{-u+\lambda_{i}+2n+1/2-i}{-u+2n+1/2-i}.
\end{equation}
For a nested subalgebra, the same determinant formula applies with its rank and highest weight. In a GT pattern these data are carried by the even row, so no intermediate odd-row label enters at this stage. On the basis vector labelled by $\Lambda$, the eigenvalue is
\begin{equation}
\mathcal{D}_{\Lambda}^{(2k)}(u)=\alpha_{2k}(u)\prod_{i=1}^{k}\frac{u+\tilde{\lambda}_{k,i}+3/2-i}{u+3/2-i}\frac{-u+\tilde{\lambda}_{k,i}+2k+1/2-i}{-u+2k+1/2-i}.\label{eq:D2k}
\end{equation}
Substituting into (\ref{eq:Ym2d}), we obtain the nested $Y^{-}(2)$ Sklyanin-determinant eigenvalues:
\begingroup\fontsize{8.5}{10.5}\selectfont
\begin{equation}
d_{\Lambda}^{(k)}(u)=\frac{\mathcal{D}_{\Lambda}^{(2k)}(u+2k-2)}{\mathcal{D}_{\Lambda}^{(2k-2)}(u+2k-2)}=\frac{1}{(u-3/2)(u-1/2)}\frac{\prod_{i=1}^{k}(u+\tilde{\lambda}_{k,i}+2k-1/2-i)(u-\tilde{\lambda}_{k,i}-5/2+i)}{\prod_{i=1}^{k-1}(u+\tilde{\lambda}_{k-1,i}+2k-1/2-i)(u-\tilde{\lambda}_{k-1,i}-1/2+i)},
\end{equation}
\endgroup
where we used
\begin{equation}
\frac{\alpha_{2k}(u+2k-2)}{\alpha_{2k-2}(u+2k-2)}=\frac{u+k-1/2}{u+k-3/2}.
\end{equation}
We next determine the G-operator eigenvalues. This requires a factorization of the nested $Y^{-}(2)$ $K$-matrices. Start with $\mathbf{k}^{(n)}$, whose entries commute with the subalgebra $Y^{-}(2n-2)$ \cite{Gombor:2025wvu}. We may therefore restrict to the space of $\mathfrak{sp}_{2n-2}$ highest-weight vectors with fixed weight $(\tilde{\lambda}_{n-1,1},\dots\tilde{\lambda}_{n-1,n-1})\equiv(\zeta_{1},\dots\zeta_{n-1})=\zeta$. Denote this invariant subspace by $V(\lambda)_{\zeta}^{+}$. By \cite[Theorem 5.2]{Molev:98}, the action of $\mathbf{k}^{(n)}$ on this space is irreducible and can be realized through a tensor product of $Y(2)$ representations:
\begin{equation}
V(\lambda)_{\zeta}^{+}\simeq L(\alpha_{1},\beta_{1}|\theta_{1})\otimes\dots\otimes L(\alpha_{n},\beta_{n}|\theta_{n}),
\end{equation}
where
\begin{align}
\alpha_{i} & =\min(\lambda_{i},\zeta_{i-1}),\quad i=2,\dots,n,\quad\alpha_{1}=\lambda_{1},\\
\beta_{i} & =\max(\lambda_{i+1},\zeta_{i}),\quad i=1,\dots,n-1,\quad\beta_{n}=0,\\
\theta_{i} & =n+1/2-i.
\end{align}
The tensor factors resolve the multiplicity space associated with two adjacent symplectic highest weights. The interlacing bounds fix their upper and lower weights, while the twisted-Yangian embedding fixes their spectral shifts. We can therefore study the remaining boundary operators through a product of rank-two evaluation factors.

The Lax operator of each evaluation factor is
\begin{equation}
L_{i,j}^{(k)}(u)=\delta_{i,j}+\frac{1}{u+\theta_{k}}E_{j,i}^{(\alpha_{k},\beta_{k})},
\end{equation}
and the associated monodromy matrix is
\begin{equation}
\mathbf{L}_{0}^{\zeta}(u)=L_{0,1}^{(1)}(u)\dots L_{0,n}^{(n)}(u).
\end{equation}
The relevant Sklyanin minor acts on the highest-weight vector as follows \cite[Proposition 4.7]{Molev:06}:
\begin{equation}
\alpha_{-2n+2}(u)\mathbf{K}_{1,2,\dots,2n-2,2n}^{1,2,\dots,2n-2,2n-1}(u+n-1)\zeta_{\Lambda_{0}}=\prod_{i=1}^{n}\frac{u-\theta_{i}-\alpha_{i}}{u-\theta_{i}}\frac{u+\theta_{i}+\beta_{i}}{u+\theta_{i}}\zeta_{\Lambda_{0}}.
\end{equation}
The tensor-product realization fixes the matrix structure of the action on the multiplicity space. To match its normalization to that of the Sklyanin minor, we introduce a scalar function $A(u)$ and write
\begin{equation}
\mathbf{K}_{1,2,\dots,2n-2,2n-2+a}^{1,2,\dots,2n-2,2n-2+b}(u+n-1)\Biggr|_{V(\lambda)_{\zeta}^{+}}=-A(u)\sum_{c,d}\mathbf{L}_{c,a}^{\zeta}(u)\epsilon_{c,d}\mathbf{L}_{d,b}^{\zeta}(-u).
\end{equation}
In particular,
\begin{equation}
\mathbf{K}_{1,2,\dots,2n-2,2n}^{1,2,\dots,2n-2,2n-1}(u+n-1)\Biggr|_{V(\lambda)_{\zeta}^{+}}=A(u)\left(\mathbf{L}_{2,2}^{\zeta}(u)\mathbf{L}_{1,1}^{\zeta}(-u)-\mathbf{L}_{1,2}^{\zeta}(u)\mathbf{L}_{2,1}^{\zeta}(-u)\right),
\end{equation}
and hence, on the highest-weight vector,
\begin{equation}
\mathbf{K}_{1,2,\dots,2n-2,2n}^{1,2,\dots,2n-2,2n-1}(u+n-1)\zeta_{\Lambda_{0}}=A(u)\prod_{i=1}^{n}\frac{u-\theta_{i}-\alpha_{i}}{u-\theta_{i}}\frac{u+\theta_{i}+\beta_{i}}{u+\theta_{i}}\zeta_{\Lambda_{0}},
\end{equation}
where we used
\begin{equation}
\mathbf{L}_{1,1}^{\zeta}(u)\zeta_{\Lambda_{0}}=\prod_{i=1}^{n}\frac{u+\theta_{i}+\alpha_{i}}{u+\theta_{i}}\zeta_{\Lambda_{0}},
\quad\mathbf{L}_{2,2}^{\zeta}(u)\zeta_{\Lambda_{0}}=\prod_{i=1}^{n}\frac{u+\theta_{i}+\beta_{i}}{u+\theta_{i}}\zeta_{\Lambda_{0}}.
\end{equation}
Comparing this expression with the highest-weight action of the Sklyanin minor, we find that the weight-dependent products coincide. The remaining scalar factor is therefore
\begin{equation}
A(u)=\frac{1}{\alpha_{-2n+2}(u)}.
\end{equation}
Using (\ref{eq:connKk}), we obtain
\begin{equation}
\mathbf{k}_{a,b}^{(n)}(u)=\left[\mathbf{D}^{(2n-2)}(u+2n-2)\right]^{-1}\mathbf{K}_{1,2,\dots,2n-2,2n-2+a}^{1,2,\dots,2n-2,2n-2+b}(u+2n-2),
\end{equation}
and therefore
\begingroup\fontsize{10}{12}\selectfont
\begin{equation}
\mathbf{k}_{a,b}^{(n)}(u)\Biggr|_{V(\lambda)_{\zeta}^{+}}=-\frac{1}{\alpha_{-2n+2}(u+n-1)\mathcal{D}_{\zeta}^{(2n-2)}(u+2n-2)}\sum_{c,d}\mathbf{L}_{c,a}^{\zeta}(u+n-1)\epsilon_{c,d}\mathbf{L}_{d,b}^{\zeta}(-u-n+1).
\end{equation}
\endgroup
The corresponding deformation is
\begingroup\fontsize{10}{12}\selectfont
\begin{multline}
\tilde{\mathbf{k}}_{a,b}^{(n)}(u)\Biggr|_{V(\lambda)_{\zeta}^{+}}=
\frac{1}{\alpha_{-2n+2}(u+n-1)\mathcal{D}_{\zeta}^{(2n-2)}(u+2n-2)} \times \\
\left((u+n-1/2)\mathbf{L}_{1,a}^{\zeta}(u+n-1)\mathbf{L}_{1,b}^{\zeta}(-u-n+1)-\sum_{c,d}\mathbf{L}_{c,a}^{\zeta}(u+n-1)\epsilon_{c,d}\mathbf{L}_{d,b}^{\zeta}(-u-n+1)\right).
\end{multline}
\endgroup
For the odd-level G-operator, we need the first diagonal entry of the deformed matrix. In the tensor-product realization, the undeformed contribution to this entry shifts the total weight, whereas the added product of diagonal monodromy entries preserves it. Ordering the weight sectors therefore makes the operator block triangular. Its eigenvalues are determined by the weight-preserving part, namely
\begin{equation}
\mathbf{O}^{\lambda,\zeta}(u)=\frac{u+n-1/2}{\alpha_{-2n+2}(u+n-1)\mathcal{D}_{\zeta}^{(2n-2)}(u+2n-2)}\mathbf{L}_{1,1}^{\zeta}(u+n-1)\mathbf{L}_{1,1}^{\zeta}(-u-n+1).
\end{equation}
The diagonal monodromy entry can in turn be triangularized in an ordered tensor-product weight basis. Its diagonal entries are products of local diagonal Lax entries. The local weight in each factor supplies one spectral label, so the eigenvalues are parametrized by $\gamma=(\gamma_{1},\dots,\gamma_{n})$:
\begingroup\fontsize{9.5}{11.5}\selectfont
\begin{equation}
\mathcal{O}_{\gamma}^{\lambda,\zeta}(u)=\frac{u+n-1/2}{\alpha_{-2n+2}(u+n-1)\mathcal{D}_{\zeta}^{(2n-2)}(u+2n-2)}\prod_{i=1}^{n}\frac{u+2n-1/2-i+\gamma_{i}}{u+2n-1/2-i}\prod_{i=1}^{n}\frac{-u+3/2-i+\gamma_{i}}{-u+3/2-i},
\end{equation}
\endgroup
where
\begin{equation}
\beta_{i}\leq\gamma_{i}\leq\alpha_{i}.
\end{equation}
Substituting the even-rank determinant eigenvalue, the scalar normalization factors cancel against the weight-independent Lax denominators. The result separates the multiplicity-space labels in the numerator from the fixed lower highest weight in the denominator:
\begin{equation}
\mathcal{O}_{\gamma}^{\lambda,\zeta}(u)=\frac{1}{u-1/2}\frac{\prod_{i=1}^{n}(u-3/2+i-\gamma_{i})(u+2n-1/2-i+\gamma_{i})}{\prod_{i=1}^{n-1}(u-1/2+i-\zeta_{i})(u+2n-1/2-i+\zeta_{i})}.
\end{equation}
Substituting the definitions of $\alpha_{i}$ and $\beta_{i}$, the allowed values satisfy
\begin{equation}
\max(\lambda_{i+1},\zeta_{i})\leq\gamma_{i}\leq\min(\lambda_{i},\zeta_{i-1}),
\end{equation}
where $\lambda_{n+1}=\zeta_n=0$ and $\zeta_0=\infty$.
These bounds identify the remaining spectral labels with the intermediate GT row. Indeed, once the rows $\tilde{\lambda}_{n,i}$ and $\tilde{\lambda}_{n-1,i}$ are fixed, the entries $\mu_{n,i}$ satisfy
\begin{equation}
\max(\tilde{\lambda}_{n,i+1},\tilde{\lambda}_{n-1,i})\leq\mu_{n,i}\leq\min(\tilde{\lambda}_{n,i},\tilde{\lambda}_{n-1,i-1}),
\end{equation}
where $\tilde{\lambda}_{n,n+1}=\tilde{\lambda}_{n-1,n}=0$ and $\tilde{\lambda}_{n-1,0}=\infty$.
With $\tilde{\lambda}_{n,i}\equiv\lambda_{i}$ and $\tilde{\lambda}_{n-1,i}\equiv\zeta_{i}$, these are precisely the bounds on $\gamma_{i}$. We can therefore identify these spectral labels with the GT entries $\mu_{n,i}$. The resulting G-operator eigenvalue is
\begin{equation}
\mathcal{G}_{\Lambda}^{(2n-1)}(u)=\frac{1}{u-1/2}\frac{\prod_{i=1}^{n}(u-3/2+i-\mu_{n,i})(u+2n-1/2-i+\mu_{n,i})}{\prod_{i=1}^{n-1}(u-1/2+i-\tilde{\lambda}_{n-1,i})(u+2n-1/2-i+\tilde{\lambda}_{n-1,i})}.\label{eq:G2nm1}
\end{equation}
Replacing $n$ by $k$ gives the eigenvalues of any odd-level operator $\mathbf{G}^{(2k-1)}$. The even-level eigenvalues of $\mathbf{G}^{(2k)}$ follow from (\ref{eq:G2k}), (\ref{eq:G2nm1}), and (\ref{eq:D2k}):
\begin{equation}
\mathcal{G}_{\Lambda}^{(2k)}(u)=\frac{1}{u-1/2}
\prod_{j=1}^{k}\frac{(u+\tilde{\lambda}_{k,j}+2k-j+1/2)(u-\tilde{\lambda}_{k,j}+j-3/2)}{(u+\mu_{k,j}+2k-j+1/2)(u-\mu_{k,j}+j-1/2)}.
\end{equation}
Taking ratios of consecutive G-eigenvalues now gives the F-eigenvalues. Each ratio introduces a neighbouring GT row, so the resulting functions depend on three adjacent rows of the pattern:
\begin{multline}
\mathcal{F}_{\Lambda}^{(2k-1)}(u)  = \frac{\prod_{j=1}^{k-1}(u+\tilde{\lambda}_{k-1,j}+2k-j-1/2)(u-\tilde{\lambda}_{k-1,j}+j-1/2)}{\prod_{j=1}^{k}(u+\mu_{k,j}+2k-j-1/2)(u-\mu_{k,j}+j-3/2)} \times  \\
  \frac{\prod_{j=1}^{k}(u+\tilde{\lambda}_{k,j}+2k-j+1/2)(u-\tilde{\lambda}_{k,j}+j-3/2)}{\prod_{j=1}^{k}(u+\mu_{k,j}+2k-j+1/2)(u-\mu_{k,j}+j-1/2)},
 \end{multline}
 and
 \begin{multline}
\mathcal{F}_{\Lambda}^{(2k)}(u)  = \frac{\prod_{j=1}^{k+1}(u+\mu_{k+1,j}+2k-j+3/2)(u-\mu_{k+1,j}+j-3/2)}{\prod_{j=1}^{k}(u+\tilde{\lambda}_{k,j}+2k-j+3/2)(u-\tilde{\lambda}_{k,j}+j-1/2)} \times \\
\frac{\prod_{j=1}^{k}(u+\mu_{k,j}+2k-j+1/2)(u-\mu_{k,j}+j-1/2)}{\prod_{j=1}^{k}(u+\tilde{\lambda}_{k,j}+2k-j+1/2)(u-\tilde{\lambda}_{k,j}+j-3/2)}.
\end{multline}
After reparametrization, they become
\begingroup\fontsize{9.5}{11.5}\selectfont
\begin{align}
\tilde{\mathcal{F}}_{\Lambda}^{(2k-1)}(u) & =  -\sqrt{\frac{u^{2}}{u^{2}+1/4}}\frac{\prod_{j=1}^{k-1}(u^{2}+(\tilde{\lambda}_{k-1,j}+k-j)^{2})}{\prod_{j=1}^{k}(u^{2}+(\mu_{k,j}+k-j)^{2})}\frac{\prod_{j=1}^{k}(u^{2}+(\tilde{\lambda}_{k,j}+k-j+1)^{2})}{\prod_{j=1}^{k}(u^{2}+(\mu_{k,j}+k-j+1)^{2})},\\
\tilde{\mathcal{F}}_{\Lambda}^{(2k)}(u) & = -\sqrt{\frac{u^{2}}{u^{2}+1/4}}\frac{\prod_{j=1}^{k}(u^{2}+(\mu_{k,j}+k-j+1/2)^{2})}{\prod_{j=1}^{k}(u^{2}+(\tilde{\lambda}_{k,j}+k-j+1/2)^{2})}\frac{\prod_{j=1}^{k+1}(u^{2}+(\mu_{k+1,j}+k-j+3/2)^{2})}{\prod_{j=1}^{k}(u^{2}+(\tilde{\lambda}_{k,j}+k-j+3/2)^{2})}.
\end{align}
\endgroup
Using the uniform notation $\lambda_{i,j}$ for the GT entries, as in (\ref{eq:lamt}), we can combine these expressions into
\begingroup\fontsize{8.8}{10.8}\selectfont
\begin{equation}
\tilde{\mathcal{F}}_{\Lambda}^{(k)}(u)= - \sqrt{\frac{u^{2}}{u^{2}+1/4}}\frac{\prod_{j=1}^{\left\lceil \frac{k-1}{2}\right\rceil }(u^{2}+(\lambda_{k-1,j}+k/2-j+1/2)^{2})}{\prod_{j=1}^{\left\lceil k/2\right\rceil }(u^{2}+(\lambda_{k,j}+k/2-j+1/2)^{2})}\frac{\prod_{j=1}^{\left\lceil \frac{k+1}{2}\right\rceil }(u^{2}+(\lambda_{k+1,j}+k/2-j+3/2)^{2})}{\prod_{j=1}^{\left\lceil k/2\right\rceil }(u^{2}+(\lambda_{k,j}+k/2-j+3/2)^{2})}.
\end{equation}
\endgroup
Finally, introducing the shifted weights
\begin{equation}
\gamma_{k,j}=\lambda_{k,j}+\frac{k}{2}-j+\frac{1}{2}
\end{equation}
gives
\begin{equation}
\tilde{\mathcal{F}}_{\Lambda}^{(k)}(u)=-\sqrt{\frac{u^{2}}{u^{2}+1/4}}\frac{\prod_{j=1}^{\left\lceil \frac{k-1}{2}\right\rceil }(u^{2}+(\gamma_{k-1,j}+1/2)^{2})}{\prod_{j=1}^{\left\lceil k/2\right\rceil }(u^{2}+\gamma_{k,j}^{2})}\frac{\prod_{j=1}^{\left\lceil \frac{k+1}{2}\right\rceil }(u^{2}+(\gamma_{k+1,j}+1/2)^{2})}{\prod_{j=1}^{\left\lceil k/2\right\rceil }(u^{2}+(\gamma_{k,j}+1)^{2})}.
\end{equation}
\section{Conclusions}\label{sec:conclusions}

We have obtained explicit overlap functions for integrable MPSs associated with arbitrary finite-dimensional irreducible evaluation representations of the orthogonal and symplectic twisted Yangians. The functions are parametrized by Gelfand--Tsetlin patterns, whose admissible entries are fixed by the highest weight of the underlying Lie-algebra representation. This gives a uniform description of entire families of boundary states without constructing explicit generator matrices or diagonalizing the associated quasi-determinants separately for each representation.

In the orthogonal case, the determinant factorization directly gives the G- and F-eigenvalues from the central elements of nested subalgebras. In the symplectic case, the singular undeformed odd-level operators require a two-step construction. Combining even-rank Sklyanin determinants with Sklyanin minors and the factorization of rank-two $K$-matrices on multiplicity spaces determines the missing odd-level spectra. Both routes lead to closely related product formulas in shifted GT weights.

These results determine the representation-dependent functions in the normalized overlap formula. The universal Gaudin-determinant factor retains the dependence on the Bethe state, while the pseudovacuum coefficients must be supplied for the chosen spin chain and MPS normalization. This separation makes the formulas applicable across different quantum spaces sharing the same boundary representation.

The orthogonal construction also expresses overlap functions beyond evaluation representations in terms of highest weights along the nested twisted-Yangian subalgebras. Extending this description requires control of the corresponding branching data. A similar treatment of more general symplectic representations is a natural further direction. More broadly, the results show how central elements of nested boundary algebras provide the spectral data needed for exact MPS overlaps.

\section*{Acknowledgments}

This paper was supported by the NKFIH grant STARTING-152796 and the János Bolyai Research Scholarship of the Hungarian Academy of Sciences.

\paragraph*{Use of artificial intelligence tools}

ChatGPT was used to assist in reviewing and finalizing the manuscript.

\appendix

\section{Conventions for the $K$-matrices}\label{app:conventions}

This appendix relates the $K$-matrix convention used in the overlap construction to the convention in which twisted-Yangian highest weights are defined. We first transform the crossed $R$-matrix and the reflection equation, then combine this transformation with the auxiliary-space change of basis introduced in Section~\ref{sec:Representations-of-the}. For evaluation representations, this gives the highest-weight functions directly in terms of the Lie-algebra weights. The final definition of the reflected highest-weight functions fixes the notation used in the Sklyanin-determinant eigenvalues.

Our convention uses
\begin{equation}
\bar{R}_{1,2}(u)=R_{1,2}^{T_{1}}(-u)=R_{1,2}^{T_{2}}(-u),
\end{equation}
where $T_{i}$ denotes ordinary transposition in the $i$th auxiliary space. An alternative convention is defined by
\begin{equation}
\bar{R}_{1,2}^{c}(u)=V_{1}\bar{R}_{1,2}(u)V_{1}^{-1},
\end{equation}
with the $N\times N$ matrix $V$ from (\ref{eq:Vdef}). It satisfies
\begin{equation}
V^{-1}=V^{T}=\pm V,
\end{equation}
where the plus and minus signs correspond to $\mathfrak{so}_{N}$ and $\mathfrak{sp}_{N}$, respectively. By the $\mathfrak{gl}_{N}$ symmetry of the $R$-matrix, the transformed matrix
\begin{equation}
\mathbf{K}^{c}(u)=V^{-1}\mathbf{K}(u)
\end{equation}
satisfies the boundary Yang--Baxter equation in the new convention:
\begin{equation}
R_{1,2}(v-u)\mathbf{K}_{1}^{c}(u)\bar{R}_{1,2}^{c}(-u-v)\mathbf{K}_{2}^{c}(v)=\mathbf{K}_{2}^{c}(v)\bar{R}_{1,2}^{c}(-u-v)\mathbf{K}_{1}^{c}(u)R_{1,2}(v-u).
\end{equation}
Here we used
\begin{equation}
V_{1}^{-1}V_{2}^{-1}R_{1,2}(u)V_{1}V_{2}=R_{1,2}(u),\quad\bar{R}_{1,2}^{c}(u)=V_{1}\bar{R}_{1,2}(u)V_{1}^{-1}=V_{2}\bar{R}_{1,2}(u)V_{2}^{-1}.
\end{equation}
The boundary Yang--Baxter equation is also covariant under the transformation
\begin{equation}
\tilde{\mathbf{K}}^{c}(u)=V^{-1}B^{T}V\mathbf{K}^{c}(u)B,
\end{equation}
for any $B\in GL(N)$. In terms of the original $K$-matrix,
\begin{equation}
\tilde{\mathbf{K}}^{c}(u)=V^{-1}B^{T}\mathbf{K}(u)B.
\end{equation}
For the change of basis defined in Section~\ref{sec:Representations-of-the},
\begin{equation}
V^{-1}B^{T}=\begin{cases}
B^{-1}, & \text{for }\mathfrak{so}_{N},\\
B^{-1}U^{-1}, & \text{for }\mathfrak{sp}_{N},
\end{cases}
\end{equation}
and hence
\begin{equation}
\tilde{\mathbf{K}}^{c}(u)=\begin{cases}
B^{-1}\mathbf{K}(u)B, & \text{for }\mathfrak{so}_{N},\\
B^{-1}U^{-1}\mathbf{K}(u)B, & \text{for }\mathfrak{sp}_{N}.
\end{cases}
\end{equation}
Substituting the evaluation homomorphisms
\begin{equation}
\mathbf{K}(u)=\begin{cases}
1+\frac{1}{u+1/2}F, & \text{for }\mathfrak{so}_{N},\\
U(1+\frac{1}{u-1/2}F) & \text{for }\mathfrak{sp}_{N}
\end{cases}
\end{equation}
gives, in the new convention,
\begin{equation}
\tilde{\mathbf{K}}^{c}(u)=1+\frac{1}{u\pm1/2}F^{c}.
\end{equation}
In this convention, a highest-weight vector in the boundary space satisfies
\begin{align}
\tilde{\mathbf{K}}_{i,j}^{c}(u)\zeta & =0,\quad1\leq i<j\leq N, \\
\tilde{\mathbf{K}}_{i,i}^{c}(u)\zeta & =\mu_{i}(u)\zeta,\quad i=1,\dots,N,\label{eq:hw}
\end{align}
where $\mu_{i}(u)$ are the highest-weight functions. For evaluation representations,
\begin{equation}
\mu_{i}(u)=\frac{u\pm1/2+\lambda_{i}}{u\pm1/2}.
\end{equation}
For more general $K$-matrices, we likewise define the highest-weight functions by (\ref{eq:hw}) in the chosen convention. To write the Sklyanin-determinant eigenvalues, it is convenient to introduce
\begin{equation}
\tilde{\mu}_{i}(u)=\begin{cases}
\mu_{N+1-i}(-u), & i=1,\dots,n,\\
\mu_{i}(u) & i=n+1,\dots,N.
\end{cases}
\end{equation}
\section{Proofs of the factorization formulas}\label{app:factorization}

This appendix proves the factorizations (\ref{eq:DDG}) and (\ref{eq:DDd}) that connect the overlap operators to nested Sklyanin determinants. We first recall how a quasi-determinant is recovered from a block of the inverse matrix. We then apply the fused definition of the Sklyanin determinant, isolating one auxiliary direction in the orthogonal case and two in the symplectic case. These two reductions explain why the orthogonal nesting produces a single G-operator, whereas the symplectic nesting produces a rank-two Sklyanin determinant.

We use a standard property of quasi-determinants. Consider an $N\times N$ matrix $X$ in block form,
\begin{equation}
X=\left(\begin{array}{cc}
A & B\\
C & D
\end{array}\right),
\end{equation}
where $A$, $B$, $C$, and $D$ are matrices over the underlying ring of sizes $M\times M$, $M\times(N-M)$, $(N-M)\times M$, and $(N-M)\times(N-M)$, respectively. Assuming that $A$ is invertible, define the block quasi-determinant by
\begin{equation}
\left(\begin{array}{cc}
A & B\\
C & \boxed{D}
\end{array}\right):=D-CA^{-1}B.
\end{equation}
Provided all displayed inverses exist, the following block-inverse identity holds over a noncommutative ring. In particular, the lower-right block of the inverse is the inverse of the Schur complement, with the order of all factors retained:
\begin{equation}
X^{-1}=\left(\begin{array}{cc}
\tilde{A} & \tilde{B}\\
\tilde{C} & \tilde{D}
\end{array}\right)=\left(\begin{array}{cc}
(A-BD^{-1}C)^{-1} & -A^{-1}B(D-CA^{-1}B)^{-1}\\
-D^{-1}C(A-BD^{-1}C)^{-1} & (D-CA^{-1}B)^{-1}
\end{array}\right).
\end{equation}
In particular,
\begin{equation}
\tilde{D}=(D-CA^{-1}B)^{-1},
\end{equation}
so the quasi-determinant can equivalently be written as
\begin{equation}
\left(\begin{array}{cc}
A & B\\
C & \boxed{D}
\end{array}\right)=\tilde{D}^{-1}.
\end{equation}
Applying this identity with $M=N-1$ gives
\begin{equation}
\left(\left(\mathbf{K}^{-1}\right)_{N,N}\right)^{-1}=\left|\begin{array}{cccc}
\mathbf{K}_{1,1} & \mathbf{K}_{1,2} & \dots & \mathbf{K}_{1,N}\\
\mathbf{K}_{2,1} & \mathbf{K}_{2,2} & \dots & \mathbf{K}_{2,N}\\
\vdots & \vdots & \ddots & \vdots\\
\mathbf{K}_{N,1} & \mathbf{K}_{N,2} & \dots & \boxed{\mathbf{K}_{N,N}}
\end{array}\right|=\mathbf{G}^{(N)},\label{eq:Gid}
\end{equation}
and with $M=N-2$ gives
\begingroup\fontsize{8.8}{10.8}\selectfont\setlength{\arraycolsep}{2pt}
\begin{equation}
\mathbf{k}^{(n)}=\left|\begin{array}{cc}
\begin{array}{ccc}
\mathbf{K}_{1,1} & \dots & \mathbf{K}_{1,2n-2}\\
\vdots & \ddots & \vdots\\
\mathbf{K}_{2n-2,1} & \dots & \mathbf{K}_{2n-2,2n-2}
\end{array} & \begin{array}{cc}
\mathbf{K}_{1,2n-1} & \mathbf{K}_{1,2n}\\
\vdots & \vdots\\
\mathbf{K}_{2n-2,2n-1} & \mathbf{K}_{2n-2,2n}
\end{array}\\
\begin{array}{ccc}
\mathbf{K}_{2n-1,1} & \dots & \mathbf{K}_{2n-1,2n-2}\\
\mathbf{K}_{2n,1} & \dots & \mathbf{K}_{2n,2n-2}
\end{array} & \boxed{\begin{array}{cc}
\mathbf{K}_{2n-1,2n-1} & \mathbf{K}_{2n-1,2n}\\
\mathbf{K}_{2n,2n-1} & \mathbf{K}_{2n,2n}
\end{array}}
\end{array}\right|=\left(\begin{array}{cc}
\left(\mathbf{K}^{-1}\right)_{2n-1,2n-1} & \left(\mathbf{K}^{-1}\right)_{2n-1,2n}\\
\left(\mathbf{K}^{-1}\right)_{2n,2n-1} & \left(\mathbf{K}^{-1}\right)_{2n,2n}
\end{array}\right)^{-1}.\label{eq:kid}
\end{equation}
\endgroup
\subsection{Proof of (\ref{eq:DDG})}

For the one-step factorization, we separate the last auxiliary direction from the first $N-1$. The inverse-block identity then turns its contribution into a single quasi-determinant. The spectral shift is inherited from the last parameter in the fused product. We begin with the definition
\begin{equation}
A_{N}\left\langle \mathbf{K}_{1},\dots,\mathbf{K}_{N}\right\rangle =A_{N}\mathrm{sdet}\mathbf{K}(u).\label{eq:sDef}
\end{equation}
The fused product on the left-hand side can be expanded as
\begin{equation}
A_{N}\left\langle \mathbf{K}_{1},\dots,\mathbf{K}_{N}\right\rangle =A_{N}\left\langle \mathbf{K}_{1},\dots,\mathbf{K}_{N-1}\right\rangle \bar{R}_{1,N}\bar{R}_{2,N}\dots\bar{R}_{N-1,N}\mathbf{K}_{N}.
\end{equation}
Substituting into the definition and rearranging, we obtain
\begin{equation}
A_{N}\left\langle \mathbf{K}_{1},\dots,\mathbf{K}_{N-1}\right\rangle \bar{R}_{1,N}\bar{R}_{2,N}\dots\bar{R}_{N-1,N}=\mathrm{sdet}\mathbf{K}(u)A_{N}\mathbf{K}_{N}^{-1}.
\end{equation}
We apply this identity to $e_{a_{1}}\otimes\dots\otimes e_{a_{N-1}}\otimes e_{N}$, where $1\leq a_{j}\leq N-1$ for all $j=1,\dots,N-1$. Using
\begin{equation}
\bar{R}_{1,N}\bar{R}_{2,N}\dots\bar{R}_{N-1,N}\left(e_{a_{1}}\otimes\dots\otimes e_{a_{N-1}}\otimes e_{N}\right)=\left(e_{a_{1}}\otimes\dots\otimes e_{a_{N-1}}\otimes e_{N}\right)
\end{equation}
and
\begingroup\fontsize{9.5}{11.5}\selectfont
\begin{equation}
A_{N}\mathbf{K}_{N}(u-N+1)^{-1}\left(e_{a_{1}}\otimes\dots\otimes e_{a_{N-1}}\otimes e_{N}\right)=\left(\mathbf{K}(u-N+1)^{-1}\right)_{N,N}A_{N}\left(e_{a_{1}}\otimes\dots\otimes e_{a_{N-1}}\otimes e_{N}\right),
\end{equation}
\endgroup
we obtain
\begingroup\fontsize{8.2}{10.2}\selectfont
\begin{equation}
A_{N}\left\langle \mathbf{K}_{1},\dots,\mathbf{K}_{N-1}\right\rangle \left(e_{a_{1}}\otimes\dots\otimes e_{a_{N-1}}\otimes e_{N}\right)=\mathrm{sdet}\mathbf{K}(u)\left(\mathbf{K}(u-N+1)^{-1}\right)_{N,N}A_{N}\left(e_{a_{1}}\otimes\dots\otimes e_{a_{N-1}}\otimes e_{N}\right).
\end{equation}
\endgroup
Summing over $1\leq a_{j}\leq N-1$ gives
\begin{equation}
A_{N-1}\left\langle \mathbf{K}_{1},\dots,\mathbf{K}_{N-1}\right\rangle \mathbf{G}^{(N)}(u-N+1)=\mathrm{sdet}\mathbf{K}(u)A_{N-1},
\end{equation}
where we used (\ref{eq:Gid}). Recognizing the Sklyanin determinant of $Y^{+}(N-1)$ on the left-hand side, we obtain
\begin{equation}
\mathrm{sdet}\mathbf{K}^{[N]}(u)=\mathrm{sdet}\mathbf{K}^{[N-1]}(u)\mathbf{G}^{(N)}(u-N+1).
\end{equation}
The same argument at each level yields (\ref{eq:DDG}).

\subsection{Proof of (\ref{eq:DDd})}

For the two-step factorization, we retain the last two auxiliary directions. Their interaction through the crossed $R$-matrix must remain explicit, since together they will form the rank-two fused product. Starting from (\ref{eq:sDef}), we expand
\begingroup\fontsize{8.2}{10.2}\selectfont
\begin{equation}
A_{N}\left\langle \mathbf{K}_{1},\dots,\mathbf{K}_{N}\right\rangle =A_{N}\left\langle \mathbf{K}_{1},\dots,\mathbf{K}_{N-2}\right\rangle \bar{R}_{1,N-1}\bar{R}_{2,N-1}\dots\bar{R}_{N-2,N-1}\bar{R}_{1,N}\bar{R}_{2,N}\dots\bar{R}_{N-2,N}\mathbf{K}_{N-1}\bar{R}_{N-1,N}\mathbf{K}_{N}.
\end{equation}
\endgroup
Rearranging, we obtain
\begingroup\fontsize{8.8}{10.8}\selectfont
\begin{equation}
A_{N}\left\langle \mathbf{K}_{1},\dots,\mathbf{K}_{N-2}\right\rangle \bar{R}_{1,N-1}\bar{R}_{2,N-1}\dots\bar{R}_{N-2,N-1}\bar{R}_{1,N}\bar{R}_{2,N}\dots\bar{R}_{N-2,N}=\mathrm{sdet}\mathbf{K}(u)A_{N}\mathbf{K}_{N}^{-1}\bar{R}_{N-1,N}^{-1}\mathbf{K}_{N-1}^{-1}.
\end{equation}
\endgroup
Using the identity
\begin{equation}
\bar{R}(u)\bar{R}(-u+N)=1,
\end{equation}
we can rewrite this as
\begin{multline}
A_{N}\left\langle \mathbf{K}_{1},\dots,\mathbf{K}_{N-2}\right\rangle \bar{R}_{1,N-1}\bar{R}_{2,N-1}\dots\bar{R}_{N-2,N-1}\bar{R}_{1,N}\bar{R}_{2,N}\dots\bar{R}_{N-2,N}=\\
\mathrm{sdet}\mathbf{K}(u)A_{N}\mathbf{K}_{N}^{-1}\bar{R}_{N-1,N}(2u-N+3)\mathbf{K}_{N-1}^{-1}.
\end{multline}
We now take the $(1,2,\dots,N-2,a_{1},a_{2})\times(1,2,\dots,N-2,b_{1},b_{2})$ matrix element to obtain
\begin{multline}
\langle1,2,\dots,N-2,a_{1},a_{2}|A_{N}\left\langle \mathbf{K}_{1},\dots,\mathbf{K}_{N-2}\right\rangle |1,2,\dots,N-2,b_{1},b_{2}\rangle=\\
\mathrm{sdet}\mathbf{K}(u)\langle1,2,\dots,N-2,a_{1},a_{2}|A_{N}\mathbf{K}_{N}^{-1}\bar{R}_{N-1,N}(2u-N+3)\mathbf{K}_{N-1}^{-1}|1,2,\dots,N-2,b_{1},b_{2}\rangle,
\end{multline}
where $a_{j}=N-1,N$ and $b_{j}=N-1,N$. Here we have used
\begingroup\fontsize{10}{12}\selectfont
\begin{equation}
\bar{R}_{1,N-1}\bar{R}_{2,N-1}\dots\bar{R}_{N-2,N-1}\bar{R}_{1,N}\bar{R}_{2,N}\dots\bar{R}_{N-2,N} |1,2,\dots,N-2,b_{1},b_{2}\rangle=
|1,2,\dots,N-2,b_{1},b_{2}\rangle.
\end{equation}
\endgroup 
It follows that
\begingroup\fontsize{8.8}{10.8}\selectfont
\begin{equation}
\mathrm{sdet}\mathbf{K}^{\left[N-2\right]}(u)\langle a_{1},a_{2}|A_{2}|b_{1},b_{2}\rangle=\mathrm{sdet}\mathbf{K}(u)\langle a_{1},a_{2}|A_{2}\mathbf{K}_{2}(u-N+1)^{-1}\bar{R}_{1,2}(2u-N+3)\mathbf{K}_{1}(u-N+2)^{-1}|b_{1},b_{2}\rangle.
\end{equation}
\endgroup
This is an operator identity on $\mathbb{C}^{2}\otimes\mathbb{C}^{2}$, with $\mathbb{C}^{2}\cong\mathrm{span}(|N-1\rangle,|N\rangle)$. Thus,
\begin{equation}
\mathrm{sdet}\mathbf{K}^{\left[N-2\right]}(u)A_{2}=\mathrm{sdet}\mathbf{K}(u)A_{2}\tilde{\mathbf{K}}_{2}(u-N+1)\bar{R}_{1,2}(2u-N+3)\tilde{\mathbf{K}}_{1}(u-N+2),
\end{equation}
where
\begin{equation}
\tilde{\mathbf{K}}=\left(\begin{array}{cc}
\left(\mathbf{K}^{-1}\right)_{N-1,N-1} & \left(\mathbf{K}^{-1}\right)_{N-1,N}\\
\left(\mathbf{K}^{-1}\right)_{N,N-1} & \left(\mathbf{K}^{-1}\right)_{N,N}
\end{array}\right).
\end{equation}
By (\ref{eq:kid}), this restricted inverse is the inverse of the rank-two Schur complement. Replacing the inverse blocks accordingly and applying the inversion relation for the crossed $R$-matrix yields
\begin{equation}
\mathrm{sdet}\mathbf{K}^{\left[N-2\right]}(u)A_{2}\mathbf{k}_{1}^{(n)}(u-N+2)\bar{R}_{1,2}(-2u+N-1)\mathbf{k}_{2}^{(n)}(u-N+1)=\mathrm{sdet}\mathbf{K}(u)A_{2}.
\end{equation}
We recognize the rank-two fused product on the left-hand side. Indeed, the Sklyanin determinant of $\mathbf{k}^{(n)}$ is defined by
\begin{equation}
A_{2}\mathbf{k}_{1}^{(n)}(u-n+1)\bar{R}_{1,2}(-2u+1)\mathbf{k}_{2}^{(n)}(u-n)=A_{2}\mathrm{sdet}\mathbf{k}^{(n)}(u-n+1).
\end{equation}
Shifting the spectral parameter gives
\begin{equation}
A_{2}\mathbf{k}_{1}^{(n)}(u-N+2)\bar{R}_{1,2}(-2u+N-1)\mathbf{k}_{2}^{(n)}(u-N+1)=A_{2}\mathrm{sdet}\mathbf{k}^{(n)}(u-N+2),
\end{equation}
and hence
\begin{equation}
\mathrm{sdet}\mathbf{K}^{\left[N\right]}(u)=\mathrm{sdet}\mathbf{K}^{\left[N-2\right]}(u)\mathrm{sdet}\mathbf{k}^{(n)}(u-N+2).
\end{equation}

\bibliography{refs_2}

\end{document}